\documentstyle[12pt]{article}
\newlength{\mathspace}
\def\np#1{ Nucl. Phys. B#1}
\def\pr#1    { Phys. Rev. D#1 }
\def\pl#1{ Phys. Lett. B#1}

\def\ijmp#1  { Int. Jour. Mod. Phys. A#1 }
\def\mpl#1   { Mod. Phys. Lett. A#1 }
\def\begineq{\begin{equation}}
\def\endeq{\end{equation}}
\def\eqabegin{\begin{eqnarray}}
\def\eqaend{\end{eqnarray}}
\def\nn{\nonumber}

\begin{document}
\baselineskip=0.7cm
\setlength{\mathspace}{2.5mm}



\begin{titlepage}

    \begin{normalsize}
     \begin{flushright}
  		 CTP-TAMU-17/98, SINP-TNP/98-12\\
                 hep-th/9805180\\
     \end{flushright}
    \end{normalsize}
    \begin{Large}
       \vspace{.4cm}
       \begin{center}
         {\bf U-duality p-Branes in Diverse Dimensions}\\ 
       \end{center}
    \end{Large}

\begin{center}
           
             \vspace{.4cm}

            J. X. L{\sc u}$^1$\footnote[2]{ E-mail address: 
            jxlu@phys.physics.tamu.edu}  and   
            Shibaji R{\sc oy}$^2$\footnote{E-mail address: 
            roy@tnp.saha.ernet.in} 

            \vspace{2mm}

            $^1${\it Center for Theoretical Physics, Physics Department}\\
            {\it Texas A \& M University, College Station, TX 77843, USA}\\

            \vspace{2mm}

        $^2${\it Saha Institute of Nuclear Physics, 1/AF 
Bidhannagar, Calcutta 700 064, India}\\

      \vspace{.6cm}

    \begin{large} ABSTRACT \end{large}
        \par
\end{center}
 \begin{small}
\ \ \ \
U-duality p-branes in toroidally compactified type II 
superstring theories in space-time dimensions $10 > D \ge 4$ can be constructed 
explicitly based on the conjectured U-duality symmetries and the corresponding
known single-charge super p-brane configurations. As concrete examples, we 
first construct explicitly the $SL(3, Z)$ superstrings and 
$SL(3, Z) \times SL(2, Z)$ 0-branes as well as their corresponding magnetic 
duals in $D = 8$. For the $SL(3, Z)$ superstrings (3-branes),  each of them is 
characterized by a triplet of integers corresponding to the electric-like 
(magnetic-like) charges associated with the three 2-form gauge potentials 
present in the theory.  For the $SL(3, Z)\times SL(2,Z)$ 0-branes (4-branes), 
each of them is labelled by a pair of triplets of integers 
corresponding to the 
electric-like (magnetic-like) charges associated with the two sets of 
three 1-form 
gauge potentials. The string (3-brane) tension and central charge are shown to 
be given by $SL(3, Z)$ invariant expressions. It is argued that when any two 
of the three integers in the integral triplet are relatively prime to each 
other, the corresponding string (3-brane) is stable and does not decay into 
multiple strings (3-branes) by a `tension gap' or `charge gap' equation. 
Similar results hold also for the 0-branes (4-branes). Alongwith the 
$SL(2, Z)$ dyonic membranes of Izquierdo et. al., these examples  provide a 
further support for the conjectured $SL(3, Z) \times SL(2, Z)$ U-duality 
symmetry in this theory. Moreover, the study of these examples along with the 
previous ones provides us a recipe for constructing the U-duality p-branes of 
various supergravity theories in diverse dimensions. Constructions for these 
U-duality p-branes are also given.  
\end{small}

\end{titlepage}
\vfil\eject

\section{Introduction}

Supergravity theories in diverse dimensions have long been known to 
possess certain non-compact global symmetry groups, i.e., the Cremmer-Julia
symmetry groups [1,2]. 
Since these theories are the long wavelength limit of
various (dimensionally reduced) string theories, the discrete subgroups
of these groups have been conjectured to be promoted to the full
quantum string theories and have been named as U-duality groups [3]. From the
string theory point of view, each of these groups usually contains a 
perturbative T-duality group [4] as well as a non-perturbative strong-weak 
duality group [5,6] as its subgroups. For example, the theory we are going to 
consider explicitly in detail the $N = 2$, $D = 8$ supergravity theory, 
which is the low energy effective
action of $T^2$-compactified type IIA/IIB\footnote[1]{In lower dimensions,
type IIA and type IIB theories are equivalent by a perturbative T-duality
symmetry. But for definiteness, we will consider compactification of
type IIA theory to $D = 8$.} string theory, has a global $SL(3, R)
\times SL(2, R)$ Cremmer-Julia symmetry. The corresponding quantum type II 
string theory in $D = 8$ has been conjectured to possess the discrete U-duality
group $SL(3, Z) \times SL(2, Z)$. Since a U-duality symmetry transforms
the string coupling constant in a non-trivial way, it interchanges the strong
and weak coupling regimes of the same theory. Thus this symmetry is by nature
non-perturbative and generally it is difficult to prove the conjecture
in the perturbative framework of string theory. However, there exist
certain BPS saturated states as classical solutions [6,7,8] in these theories
whose masses and the charges do not receive any quantum corrections
due to some non-renormalization theorems of the underlying supersymmetric
theories. Thus these states are very useful to identify the non-perturbative
symmetry group of the quantum string theory.

Given a U-duality symmetry for a particular system, it is clearly artificial to 
consider p-brane solutions carrying either electric or magnetic charges 
associated with only a single $(p + 2)$-form field strength unless this field 
strength is a singlet under the U-duality symmetry, as pointed out in [9]. 
In general, we expect that there is an infinite family of such solutions 
forming  U-duality multiplets. In this paper, we first construct explicitly 
the $SL(3, Z)$ BPS saturated string-like and  3-brane-like (the magnetic dual 
of string) solutions, and the $SL(3, Z)\times SL(2,Z)$ BPS saturated 
particle-like and 4-brane-like (the magnetic dual of 0-brane) solutions by 
using the symmetry of the toroidally compactified type II string theory in 
$D = 8$.  These particular 
constructions, combined with the previous studies [9,10,11,12],  
provide us a recipe 
for constructing the general U-duality  p-brane solutions in diverse dimensions.
We then apply this recipe to construct all U-duality p-brane solutions, 
preserving half of the spacetime supersymmetry, of various supergravity 
theories 
in diverse dimensions. In exactly the same fashion, U-duality p-brane solutions
preserving less than half of spacetime supersymmetry [13] can also be 
constructed.   
The key to all such constructions of U-duality p-brane solutions is the
scalar matrices each of which parametrizes the corresponding Cremmer-Julia 
coset $G/H$ in various supergravity theories.

The $SL(3,Z)$ superstring and super 3-brane solutions as well as  the 
$SL(3, Z) \times SL(2, Z)$ superparticle and super 4-brane solutions in $D = 8$
are in a sense complementary to the dyonic membrane solutions of
Izquierdo et al [12]. The membranes are associated with a 
doublet representation of
 the $SL(2, Z)$ T-duality group and are inert under the $SL(3, Z)$ group. 
On the other
hand, the strings as well as the 3-branes are associated with a triplet 
representation of the strong-weak $SL(3, Z)$ duality group 
and are inert under the T-duality $SL(2, Z)$ group. The particles or 0-branes 
as well as the 4-branes are, however, associated with both a triplet 
representation of the $SL(3,Z)$ and a doublet representation of the 
$SL(2, Z)$ groups. Combined with their magnetic duals,
the $SL(3, Z)$ strings and $SL(3, Z)\times SL(2, Z)$ 0-branes along with the
$SL(2, Z)$ dyonic membranes of Izquierdo et. al. complete all the U-duality 
p-branes in $D = 8$.  
Therefore, the $SL(3, Z)$  string-like
solutions, the $SL(3, Z)\times SL(2,Z)$ 0-brane solutions along with 
their magnetic duals and the dyonic membrane solutions  provide a strong 
support in favor of the conjectured U-duality
symmetry in $D = 8$ quantum type II string theory. The string (3-brane) 
solutions are 
characterized by a triplet of integers corresponding to the electric-like
(magnetic-like)
charges associated with the three two-form gauge fields (one from the
NSNS sector and the other two from the RR sector) present in the theory. The
0-brane (4-brane) solutions, on the other hand, are characterized by 
a pair of triplets of integers 
corresponding to the electric-like (magnetic-like) charges associated with 
the two sets of three 1-form
gauge potentials (one set of 1-form gauge potentials has Kaluza-Klein origin
and the other set has its origin in the dimensional reductions of the 
antisymmetric 
tensors in higher dimensions). 
We will show that both the string (3-brane) tension and central charge 
associated with a general string (3-brane) solution are 
given by $SL(3, Z)$ invariant expressions. The mass and the 
central charge associated with a general 0-brane (4-brane) solution are 
given by $SL(3, Z)\times SL(2, Z)$ invariant expressions. As stated
earlier, these physical quantities remain unrenormalized in the full quantum 
theory and therefore provide a strong indication that $SL(3, Z)\times SL(2,Z)$
 is indeed a symmetry of $D = 8$ theory. 
We will also show that when any of the three pairs of integers are coprime, 
the corresponding string (3-brane) is stable as it is
prevented from decaying into multiple strings (3-branes) by a `tension gap' 
or `charge gap' equation. Similar conclusions hold also for the 0-branes and 
4-branes.  This is actually
true for all U-duality p-branes of supergravity theories in diverse dimensions.

We organize the remaining sections of this paper as follows: 
In section 2, we give a brief 
discussion of $D = 8$ NSNS strings which will provide a starting point for 
the construction of $SL(3, Z)$ strings. We demonstrate in section 3, 
using the $D = 8$ maximal supergravity as an example, how to write 
a dimensionally reduced 
bosonic action in a manifest Cremmer-Julia symmetry invariant form if this 
symmetry is realized at the level of action. This process also determines 
how the various fields transform under the underlying Cremmer-Julia symmetry
and the corresponding scalar coset matrix which are all important for the 
construction of U-duality p-branes. In particular, we provide a way to determine
the scalar coset matrix when the underlying Cremmer-Julia symmetry is not
realized at the level of action but at the level of equations of motion. 
Based on
the discussion given in section 3, we give a detail construction of the 
$SL(3, Z)$ strings in section 4. Various properties of these strings are 
discussed and the construction of the corresponding magnetic dual 
$SL (3, Z)$ 3-brane solutions are 
also given. In section 5, we construct the  
$SL(3, Z)\times SL(2, Z)$ 0-branes and the magnetic dual 4-branes, 
which completes the 
construction of all U-duality p-branes in $D = 8$ type II theory. 
Our final section consists 
of the construction of U-duality p-branes of various supergravity
theories in diverse 
dimensions.  

\section{NSNS Strings: A Brief Review}

Since we will make use of the NSNS string solution of Dabholkar et al [7] in 
$D = 8$, let us give a brief discussion of this solution. 
The low energy bosonic action common to all string theories in $D = 8$
 has the form:
\begineq
S_8 = \int\,d^8 x \sqrt {-g}\left[R - 
\frac{1}{2} \nabla_\mu \Phi \nabla^\mu \Phi - \frac{1}{12}
e^{- 2\Phi/\sqrt{3}} {\tilde F}_{\mu\nu \lambda}^{(1)} 
{\tilde F}^{(1)\,\mu\nu\lambda}\right]
\endeq
Here $g = {\rm det}\,(g_{\mu\nu})$, $g_{\mu\nu}$ being the canonical
metric which is related to the eight dimensional string metric by 
$G_{\mu\nu} = e^{\Phi/\sqrt{3}} g_{\mu\nu}$. $R$ is the scalar curvature 
with respect to the canonical
metric, $\Phi$ is the eight dimensional dilaton and 
${\tilde F}_{\mu\nu\lambda}^{(1)}$ is the field
strength associated with the Kalb-Ramond antisymmetric tensor field
$A_{\mu\nu}^{(1)}$.  The equations of motion following from (1)
admit a two parameter family of black string solution as given below [8]:
\eqabegin
ds^2 &=& -\left(1 - \frac{r_+^4}{r^4}\right)\left(1 - \frac{r_-^4}
{r^4}\right)^{-1/3} dt^2 + \left(1 - \frac{r_-^4}{r^4}\right)^{2/3}
(dx^1)^2\nn\\
& &\qquad\qquad+\left(1 - \frac{r_+^4}{r^4}\right)^{-1} \left(1 - \frac{
r_-^4}{r^4}\right)^{-5/6} dr^2 + r^2 \left(1 - \frac{r_-^4}{r^4}\right)^
{1/6} d\Omega_5^2\nn\\
e^{2\Phi} &=& \left(1 - \frac{r_-^4}{r^4}\right)^{2/\sqrt{3}},\qquad  
{\tilde F}_3^{(1)}\,\,\, =\,\,\,4\left(r_+ r_-\right)^2 
\ast e^{2\Phi/\sqrt{3}} \epsilon_5
\eqaend
Here $d\Omega_5^2$ is the metric on the unit 5-dimensional sphere and
$\epsilon_5$ is the corresponding volume form. The `$\ast$' denotes the
Hodge dual operation. $r_+$ and $r_-$ are the two parameters 
representing the two horizons with $r_+ \geq r_-$ and are related
to the charge and the mass of the black string solution. 
In the extremal limit when
$r_+ = r_-$, the solution becomes supersymmetric saturating the BPS
condition. By introducing the isotropic coordinate $\rho^4 = r^4 - r_-^4$,
the solution in the extremal limit can be written as:
\eqabegin
ds^2 &=& \left(1 + \frac{Q}{4 \rho^4}\right)^{-2/3}\left[-(dt)^2 +
(dx^1)^2\right]
+\left(1 + \frac{Q}{4\rho^4}\right)^{1/3}\left(d\rho^2 + \rho^2 d
\Omega_5^2\right)\nn\\
e^{-2\Phi} &=& \left(1 + \frac{Q}{4\rho^4}\right)^{2/\sqrt{3}}\,\,\,=\,\,\
A(\rho), \qquad {\tilde F}_3^{(1)}\,\,\,=\,\,\, Q  A^{-1/{\sqrt 3}}\,(\rho)\ast
\epsilon_5,
\eqaend 
where $Q = 4 r_-^4$. Eq.(3) represents precisely the string solution
constructed by Dabholkar et al [7] and we notice that this solution
is the extremal limit of the black string solution 
of ref.[8] in a new coordinate. $Q$ in Eq.(3) is the electric charge 
associated with the
gauge field $A_{\mu\nu}^{(1)}$ and is defined as $Q = \frac{1}{\pi^3}
\int_{S^5} \ast e^{- 2\Phi/\sqrt{3}} {\tilde F}_3^{(1)}$. Note that this 
charge is
quantized in some basic units since there also exists magnetically 
charged 3-brane solution in this theory [6,14]. It should be 
remarked here that
the solution (3) has a singularity at $\rho = 0$ since the volume
of the 5-sphere vanishes and the curvature blows up 
at that point [6]. So, the string solution (3) has been been obtained
by coupling the supergravity action (1) to a macroscopic string
source. This type of solution are usually called the `fundamental' 
solution.

As we will see, the action (1) can be regarded as a special
case of the low energy effective action of type II string theory
in $D = 8$, when some of the fields are set to zero. So, it is clear
that a more general string-like solution than that in (3) exists
when we consider the full type II theory. These general solutions
can be obtained from (3) by using the symmetry of the type II theory
in $D = 8$ as we will show.

\section{$SL(3,R)$ Invariant Action}

In order to obtain the general p-brane solution the most important
object we need is the scalar coset matrix consisting of the scalars
of the theory which parametrize the Cremmer-Julia symmetry group modded
out by its maximal compact subgroup. One way to obtain this matrix
has been outlined in ref.[15]. In this section, we will first show how
to write the low energy effective action of $D = 8$ type II theory in
$SL(3,R)$ invariant form. This process, in turn, will provide us another way
of obtaining the scalar coset matrix in this theory. This method applies 
in general whenever the Cremmer-Julia symmetry is realized at the level of
action. We will show the detail construction of this matrix below.

   The type II theory in $D = 8$ can be obtained by a $T^2$ compactification 
of $D = 10$ type IIA supergravity theory consisting of a graviton ($g_{MN}$), 
a dilaton ($\phi$) and a 2-form potential ($B_{MN}$) in its NSNS sector 
and a 1-form gauge potential (${\cal A}_M$) and a 3-form gauge potential 
($A_{MNP}$) in its RR sector. 
As discussed in detail in ref.[16], the toroidally compactified type IIA 
supergravity theories 
in $D \le 9$ can be obtained in general either from the ten dimensional 
type IIA theory or from the eleven dimensional supergravity by a set of 
successive 1-step Kaluza-Klein reductions on circles. The same procedure 
can also
be applied to the toroidal compactification of the type IIB supergravity.
In each reduction step from $(D + 1)$ to $D$ dimensions, 
the metric in $(D + 1)$ will give rise to a metric,
a Kaluza-Klein vector potential ${\cal A}_\mu$, and a ``dilatonic" scalar field 
$\varphi$ in $D$ dimensions. An $n$-index gauge potential in $(D + 1)$ 
dimensions will give rise to an $n$-index gauge potential and an 
$(n - 1)$-index gauge potential in $D$ dimensions. Following the type IIA 
reduction route to $D = 8$, we have the following bosonic 
field content: the metric $g_{\mu\nu}$, the ten dimensional type IIA dilaton 
$\phi$ together with two additional dilatonic scalars $\varphi_1$ and 
$\varphi_2$, one 3-form gauge potential $A_3$, three 2-form gauge potentials
 $A^{(i)}_2$ with $A^{(1)}_{\mu\nu} \sim B_{\mu\nu}, A^{(2)}_{\mu\nu} 
\sim A_{\mu\nu 9}, A^{(3)}_{\mu\nu} \sim A_{\mu\nu 8}$, 
three 1-form gauge potentials $A^{(i)}_1$ with $A^{(1)}_\mu \sim A_{\mu 8 9}, 
A^{(2)}_\mu \sim - B_{\mu 8}, A^{(3)}_\mu \sim B_{\mu 9}$ and another three 
1-form gauge potentials ${\cal A}^{(i)}_1$ (which can be interpreted as 
having Kaluza-Klein origin) with ${\cal A}^{(1)}_\mu \sim {\cal A}_\mu, 
{\cal A}^{(2)}_\mu \sim g_{\mu 9}/g_{99}, {\cal A}^{(3)}_\mu 
\sim g_{\mu 8}/g_{88}$, and four
0-forms (axions) $\chi_1 \sim - g_{89}/g_{88}, 
\chi_2 \sim - {\cal A}_8, \chi_3 \sim - {\cal A}_9, \rho \sim B_{89}$. 
We have used the notation such that the 
origin of the various fields can be understood from the type IIA theory 
in $D = 10$. The corresponding Lagrangian using our notation is
\eqabegin
{\cal L} &=&e \,\Bigg\{R- \frac{1}{2} 
 \,\left[(\partial \phi)^2 + (\partial \varphi_1)^2 +
(\partial \varphi_2)^2 \right] - \frac{1}{2}\,  
e^{-\phi - \frac{3}{\sqrt{7}}\varphi_1
- 2{\sqrt{\frac{3}{7}}} \varphi_2} (\partial \rho)^2 \nn\\
&\,&-\frac{1}{2}\,  \left[ e^{\frac{4}{\sqrt{7}}\varphi_1 - 2 
{\sqrt{\frac{3}{7}}} \varphi_2} (\partial \chi_1)^2 + 
e^{\frac{3}{2} \phi + \frac{1}{2\sqrt{7}}
\varphi_1 - 2{\sqrt{\frac{3}{7}}}\varphi_2} (\partial \chi_2 + \chi_1 \partial
\chi_3)^2 + e^{\frac{3}{2}\phi - \frac{\sqrt{7}}{2}\varphi_1} 
(\partial \chi_3)^2\right]\nn\\
&\,&-\frac{1}{12}\,  \left[e^{- \phi + \frac{1}{\sqrt{7}}\varphi_1 + 
\frac{2}{\sqrt{21}} 
\varphi_2} (F_3^{(1)})^2 + e^{\frac{1}{2} \phi - \frac{5}{2\sqrt{7}}\varphi_1
+\frac{2}{\sqrt{21}} \varphi_2} (F_3^{(2)})^2 + e^{\frac{1}{2}\phi + 
\frac{3}{2\sqrt{7}}\varphi_1 - \frac{4}{\sqrt{21}}\varphi_2} (F_3^{(3)})^2
\right]\nn\\
&\,&-\frac{1}{4} \, \left[e^{\frac{1}{2}\phi - \frac{5}{2\sqrt{7}}\varphi_1 - 
\frac{5}{\sqrt{21}}\varphi_2} (F_2^{(1)})^2 + e^{-\phi + \frac{1}{\sqrt{7}}
\varphi_1 - \frac{5}{\sqrt{21}}\varphi_2} (F_2^{(2)})^2 + e^{-\phi - 
\frac{3}{\sqrt{7}}\varphi_1 + \frac{1}{\sqrt{21}} \varphi_2} (F_2^{(3)})^2
\right]\nn\\
&\,&-\frac{1}{4}\,\left[ e^{\frac{3}{2}\phi + \frac{1}{2\sqrt{7}}\varphi_1 +
\frac{1}{\sqrt{21}}\varphi_2} ({\cal F}_2^{(1)})^2 + e^{\frac{4}{\sqrt{7}}
\varphi_1 + \frac{1}{\sqrt{21}}\varphi_2} ({\cal F}_2^{(2)})^2 + e^{\sqrt
{\frac{7}{3}}\varphi_2} ({\cal F}_2^{(3)})^2\right]\nn\\
&\,&-\frac{1}{48} \, e^{\frac{1}{2}\phi + \frac{3}{2\sqrt{7}}\varphi_1 +
\sqrt{\frac{3}{7}}\varphi_2} F_4^2 \Bigg\} + \frac{1}{2}\, \rho\, 
{\tilde F}_4 \wedge {\tilde F}_4\nn\\
&\,&- \frac{1}{6}\, {\tilde F}_3^{(i)}\wedge {\tilde F}_3^{(j)} 
\wedge A_2^{(k)} 
\epsilon_{ijk} - {\tilde F}_4 \wedge {\tilde F}_3^{(i)}\wedge A_1^{(i)},
\eqaend
where we have defined $e = \sqrt{- g}$, $i, j, k = 1, 2, 3$ and 
$\epsilon_{ijk}$
is totally antisymmetric with $\epsilon_{123} = 1$. We follow the notation 
in ref.[15] that field strengths without tildes include the 
various Chern-Simons modifications, 
while field strengths written with tildes do not include the modifications, 
i.e., ${\tilde F}_n^{(i)} = d A_{n -1}^{(i)}$.
The expressions for field strengths without tildes are complicated and 
given in the appendix. As in ref.[15] the wedge product is defined, 
for example,
as ${\tilde F}_4 \wedge {\tilde F}^{(i)}_3 \wedge A_1^{(i)} = \frac{1}
{4!} \frac{1}{3!} \epsilon^{\mu_1 \ldots \mu_8} {\tilde F}_{\mu_1 \ldots
\mu_4} {\tilde F}^{(i)}_{\mu_5 \mu_6 \mu_7} A^{(i)}_{\mu_8}$.

Alternatively, the same $D = 8$ supergravity can also be obtained 
from the dimensional reduction of the type IIB supergravity in $D = 10$.
Actually, it is more convenient to identify the underlying Cremmer-Julia 
symmetry $SL(3,R)\times SL(2,R)$ if we choose the basis of dilatonic scalars
that corresponds to the type IIB reduction route. Moreover, as we will see, 
one of the advantages in choosing the type IIB basis is that the $SL(2, Z)$ is 
easily understood as a T duality symmetry since its transformation does not 
involve the ten dimensional type IIB dilaton at all while the $SL(3, Z)$ is 
indeed a strong-weak duality symmetry since it contains transformations 
changing the sign of the dilaton. 
We therefore choose to work in the type IIB basis from now on. 

  The type IIA and type IIB reduction routes result in two formulations of
the $D = 8$ theory that are related to each other by the following 
orthogonal field 
redefinitions of the dilatonic scalars $\phi$ and $\varphi_1$ as
\begineq
\left(\begin{array}{c}\phi\\
      \varphi_1\end{array}\right)_{IIA} 
= \left(\begin{array}{cc} \frac{3}{4}&-\frac{\sqrt{7}}{4}\\
  			-\frac{\sqrt{7}}{4}&-\frac{3}{4}\end{array}
  			\right) 
\left(\begin{array}{c}\phi\\
  			\varphi_1\end{array}\right)_{IIB},
\endeq
which corresponds to a T-duality transformation. We can therefore obtain the 
type IIB basis Lagrangian by applying the above relation to the Lagrangian (4).
However, before we do so, we need to perform some field redefinitions which 
will greatly simplify the expressions for the field strengths without tildes 
given in the appendix and will facilitate the construction for the scalar 
coset matrix mentioned at the outset of this section.

We first perform the field redefinition 
${\cal A}_1^{(2)} \rightarrow {\cal A}_1^{(2)} + \chi_1 {\cal A}_1^{(3)}$ 
and after
that we perform  
${\cal A}_1^{(1)} \rightarrow {\cal A}_1^{(1)} + \chi_3 {\cal A}_1^{(2)} + 
\chi_2 {\cal A}_1^{(3)}$. Now we have,
\eqabegin
{\cal F}_2^{(1)} &=& {\tilde{\cal F}}_2^{(1)} - \chi_3 {\tilde{\cal F}}_2^{(2)}
- \chi_2 {\tilde{\cal F}}_2^{(3)},\nn\\
{\cal F}_2^{(2)} &=& {\tilde{\cal F}}_2^{(2)} - \chi_1 {\tilde{\cal F}}_2^{(3)},
\nn\\
{\cal F}_2^{(3)}&=& {\tilde{\cal F}}_2^{(3)}.
\eqaend
If we introduce a column vector ${\cal F}_2$ for the above three 2-form field 
strengths without tildes and a column vector ${\tilde {\cal F}}_2$ for the three
field strengths with tildes, we can write the above three equations in a 
compact form as
\begineq
{\cal F}_2 = \lambda_1 {\tilde{\cal F}}_2,
\endeq
where the matrix $\lambda_1$ is
\begineq
\lambda_1 = \left(\begin{array}{ccc}
                  1&-\chi_3&-\chi_2\\
                  0&1&-\chi_1\\
                  0&0&1 \end{array}\right).
\endeq

With the above redefinitions for ${\cal A}_1^{(2)}$ and ${\cal A}_1^{(1)}$, 
we can further perform the field redefinitions for the other three 1-form gauge
potentials as $A_1^{(i)} \rightarrow A_1^{(i)} - \rho\, {\cal A}_1^{(i)}$, the
 three 2-form gauge potentials as 
$A_2^{(i)} \rightarrow A_2^{(i)} - \frac{1}{2} \rho\, {\cal A}_1^{(j)}\wedge
{\cal A}_1^{(k)} \epsilon_{ijk}$ and the 3-form potential as
$A_3 \rightarrow A_3 - \rho\, {\cal A}_1^{(1)}\wedge {\cal A}_1^{(2)}\wedge 
{\cal A}_1^{(3)}$. The 2-form field strengths without tildes can then be
written as, 
\eqabegin
F_2^{(1)} &=& {\tilde F}_2^{(1)} - \chi_3 {\tilde F}_2^{(2)} - 
\chi_2 {\tilde F}_2^{(3)} + \rho\, ({\tilde{\cal F}}_2^{(1)} - \chi_3
{\tilde{\cal F}}_2^{(2)} - \chi_2 {\tilde{\cal F}}_2^{(3)}),\nn\\
F_2^{(2)} &=& {\tilde{\cal F}}_2^{(2)} - \chi_1 {\tilde{\cal F}}_2^{(3)} +
  		\rho\, ({\tilde{\cal F}}_2^{(2)} - \chi_1 
{\tilde{\cal F}}_2^{(3)}),\nn\\
F_2^{(3)} &=& {\tilde F}_2^{(3)} + \rho\, {\tilde{\cal F}}_2^{(3)},
\eqaend
Similarly, the three 3-form field strengths without tildes are,
\eqabegin
F_3^{(1)} &=& {\tilde F}_3^{(1)}  - \left(
{\tilde F}_2^{(2)}\wedge {\cal A}_1^{(3)} -  {\tilde F}_2^{(3)}\wedge 
{\cal A}_1^{(2)}\right),\nn\\
F_3^{(2)}&=&  \chi_3 {\tilde F}_3^{(1)} + {\tilde F}_3^{(2)} -
({\tilde F}_2^{(3)}\wedge {\cal A}_1^{(1)} - {\tilde F}_2^{(1)}\wedge 
{\cal A}_1^{(3)}) - \chi_3 ({\tilde F}_2^{(2)}\wedge {\cal A}_1^{(3)} - 
{\tilde F}_2^{(3)}\wedge {\cal A}_1^{(2)}),\nn\\
F_3^{(3)} &=&(\chi_2 + \chi_1 \chi_3) {\tilde F}_2^{(1)} + \chi_1 
{\tilde F}_2^{(2)} + {\tilde F}_3^{(3)}  - 
(\chi_2 + \chi_1 \chi_3) ({\tilde F}_2^{(2)}\wedge {\cal A}_1^{(3)} - 
{\tilde F}_2^{(3)}\wedge {\cal A}_1^{(2)})\nn\\
&\quad & - \chi_1 ({\tilde F}_2^{(3)}\wedge {\cal A}_1^{(1)} -
{\tilde F}_2^{(1)}\wedge {\cal A}_1^{(3)}) - ({\tilde F}_2^{(1)}\wedge 
{\cal A}_1^{(2)} - {\tilde F}_2^{(2)}\wedge {\cal A}_1^{(2)}),
\eqaend
and the 4-form field strength without tilde is given as,
\begineq
F_4 = {\tilde F}_4 - {\tilde F}_3^{(i)}\wedge {\cal A}_1^{(i)} + \frac{1}{2}
\epsilon_{ijk} {\tilde F}_2^{(i)}\wedge {\cal A}_1^{(j)}\wedge{\cal A}_1^{(k)}.
\endeq
We can also write (9) and (10) in compact forms if the 
corresponding column vectors
are introduced, respectively, i.e., 
\begineq
F_2 = \lambda_1 ({\tilde F}_2 + \rho\, {\tilde{\cal F}}_2),
\endeq
\begineq
F_3 = \lambda_2 ({\tilde F}_3 - G_3),
\endeq
and
\begineq
F_4 = {\tilde F}_4 - {\tilde F}_3^T \wedge {\cal A}_1 + \frac{1}{2} 
\epsilon_{ijk} {\tilde F}_2^{(i)}\wedge {\cal A}_1^{(j)}\wedge{\cal A}_1^{(k)}.
\endeq
In the above, the matrix $\lambda_1$ is given by Eq.(8) and the matrix 
$\lambda_2$ is
\begineq
\lambda_2 = \left(\begin{array}{ccc}
                 1&0&0\\
                 \chi_3 &1&0\\
                 (\chi_2 + \chi_1 \chi_3)& \chi_1 &1\end{array}\right).
\endeq
Also, the components of the column vector $G_3$ are
$G_3^{(i)} \equiv \epsilon_{ijk} {\tilde F}_2^{(j)}\wedge {\cal A}_1^{(k)}$ and
`$T$' denotes the transposition.

Using the above field redefinitions the Lagrangian (4)  can be written in the
type IIB basis as follows,
\eqabegin
{\cal L} &=& e\,R - \frac{1}{2} \,e\, \left[(\partial \sigma)^2 + e^{ 2\sigma} 
(\partial \rho)^2\right] - \frac{1}{48}\, e\, 
e^{- \sigma} F_4^2 + \frac{1}{2} \rho\, F_4 \wedge F_4\nn\\
&\quad &-\frac{1}{2}\, e\, \left[(\partial \phi)^2 + (\partial \varphi)^2 + 
e^{- \phi - \sqrt{3} \varphi} (\partial \chi_1)^2 + e^{\phi - \sqrt{3} \varphi}
(\partial \chi_2 + \chi_1 \partial \chi_3)^2 + e^{ 2\phi} (\partial \chi_3)^2
\right]\nn\\
&\quad &-\frac{1}{12}\, e\,\left[ e^{- \phi + \frac{1}{\sqrt 3} \varphi} 
(F_3^{(1)})^2 + e^{\phi + \frac{1}{\sqrt 3} \varphi} (F_3^{(2)})^2 + 
e^{- \frac{2}{\sqrt 3} \varphi} (F_3^{(3)})^2 \right] - \frac{1}{6} 
\,{\tilde F}_3^{(i)}\wedge {\tilde F}_3^{(j)}\wedge A_2^{(k)} \epsilon_{ijk}
\nn\\
&\quad &-\frac{1}{4}\, e\, e^\sigma \,\left[ e^{\phi - \frac{1}{\sqrt 3} 
\varphi} (F_2^{(1)})^2 + e^{-\phi - \frac{1}{\sqrt 3} \varphi} (F_2^{(2)})^2 + 
e^{\frac{2}{\sqrt 3} \varphi} (F_2^{(3)})^2\right]\nn\\
&\quad &-\frac{1}{4}\, e\, e^{-\sigma}\,  \left[ e^{\phi - \frac{1}{\sqrt 3} 
\varphi} ({\cal F}_2^{(1)})^2 + e^{-\phi - \frac{1}{\sqrt 3} \varphi} 
({\cal F}_2^{(2)})^2 + e^{\frac{2}{\sqrt 3} \varphi} ({\cal F}_2^{(3)})^2
\right]\nn\\
&\quad & - {\tilde F}_4\wedge{\tilde F}_3^{(i)}\wedge A_1^{(i)}.
\eqaend
In obtaining the above Lagrangian, we have further made the following field 
redefinitions
\eqabegin
\varphi_1 &=& \sqrt{\frac{3}{7}} \varphi + \frac{2}{\sqrt 7} \sigma,\nn\\
\varphi_2 &=&  \frac{2}{\sqrt 7} \varphi - \sqrt{\frac{3}{7}} \sigma,
\eqaend
and have dropped surface terms.

We now re-express the above action in a manifestly $SL(3, R)$ 
invariant form
using the compact forms for various field strengths given in Eqs.(7)
and (12)--(14). This process 
also determines the scalar matrix parametrizing the coset $SL(3,R)/SO(3)$. Let 
us demonstrate how to achieve this by the following example. The kinetic terms 
for the 3-form field strengths in the above Lagrangian can be re-expressed as
\begineq
 -\frac{1}{12}e\, F_3^T \left(\begin{array}{ccc}
                               e^{- \phi + \frac{1}{\sqrt 3} \varphi}&0&0\\
  				0&e^{\phi + \frac{1}{\sqrt 3} \varphi}&0\\
  				0&0&e^{- \frac{2}{\sqrt 3} \varphi}\end{array}
\right) F_3 = -\frac{1}{12} e\, 
({\tilde F}_3 - G_3)^T {\cal M}_3 ({\tilde F}_3 - G_3),
\endeq
where Eq.(13) has been used and the scalar matrix ${\cal M}_3$ parametrizing the 
coset $SL(3,R)/SO(3)$ is
\eqabegin
{\cal M}_3 &=& \lambda_2^T \left(\begin{array}{ccc}
                                     e^{-\phi + \frac{1}{\sqrt 3} \varphi}&0&0\\
  				0&e^{\phi + \frac{1}{\sqrt 3} \varphi}&0\\
  				0&0&e^{-\frac{2}{\sqrt 3} \varphi}\end{array}
\right)\lambda_2,\nn\\
&=&e^{ \frac{\varphi}{\sqrt 3}}\left(\begin{array}{ccc}
e^{-\phi} + \chi_3^2 e^{\phi}  &  \chi_3 e^{\phi}  
&(\chi_2 + \chi_1 \chi_3)e^{-{\sqrt 3} \varphi}\\
+ (\chi_2 + \chi_1 \chi_3)^2 e^{-{\sqrt 3} \varphi}&+\chi_1 (\chi_2 + 
\chi_1 \chi_3) e^{-{\sqrt 3} \varphi}&\\
&&\\
\chi_3 e^{\phi} & e^{\phi} + \chi_1^2 e^{-{\sqrt 3} \varphi}&
\chi_1 e^{-{\sqrt 3} \varphi}\\
+ \chi_1 (\chi_2 + \chi_1 \chi_3)e^{-{\sqrt 3} \varphi}&&\\
&&\\
(\chi_2 + \chi_1 \chi_3)e^{-{\sqrt 3} \varphi}&\chi_1 e^{-{\sqrt 3} \varphi}
&e^{-{\sqrt 3} \varphi}\end{array}\right).
\eqaend
A different but equivalent form of this scalar coset matrix has been 
given in ref.[17]. Now
the same procedure can be applied to the kinetic terms for the 2-form field 
strengths $F_2^{(i)}$ and ${\cal F}_2^{(i)}$, respectively. Using this
technique we end up with the 
following compact form for the Lagrangian of $D = 8$ type II theory,
\eqabegin
{\cal L} &=& e\, R + \frac{1}{4} e\,{\rm tr} \nabla_\mu {\cal M}_2 
\nabla^\mu {\cal M}_2^{- 1} - \frac{1}{48} e\, e^{-\sigma} F_4^2 + \frac{1}{2} 
\rho\, F_4\wedge F_4\nn\\
&\quad &+ \frac{1}{4} e\,{\rm tr}\nabla_\mu {\cal M}_3 \nabla^\mu 
{\cal M}_3^{ - 1} - \frac{1}{12} e\, ({\tilde F}_3 - G_3)^T 
{\cal M}_3 ({\tilde F}_3 - G_3) - 
\frac{1}{6} {\tilde F}_3^{(i)}\wedge {\tilde F}_3^{(j)}\wedge A_2^{(k)} 
\epsilon_{ijk}\nn\\
&\quad &-\frac{1}{4} e\, e^\sigma ({\tilde F}_2 + \rho {\tilde{\cal F}}_2)^T 
{\cal M}_3^{-1} ({\tilde F}_2 + \rho {\tilde{\cal F}}_2) - \frac{1}{4} e\, 
e^{-\sigma} {\tilde{\cal F}}_2^T {\cal M}_3^{- 1} {\tilde{\cal F}}_2 - 
{\tilde F}_4\wedge {\tilde F}_3^T\wedge A_1,
\eqaend
where the scalar matrix ${\cal M}_2$  parametrizes the coset $SL(2,R)/SO(2)$ 
and is given as,
\begineq
{\cal M}_2 = e^{\sigma} \left(\begin{array}{cc}
e^{-2\sigma} + \rho^2&\rho\\
\rho&1\end{array}\right).
\endeq 
It is not difficult to check that the above action is  invariant 
under the following $SL(3,R)$ transformations:
\eqabegin
&&g_{\mu\nu} \rightarrow g_{\mu\nu},\qquad F_4 \rightarrow F_4,\qquad 
{\cal M}_2 \rightarrow {\cal M}_2,\nn\\
&&{\cal M}_3 \rightarrow \Lambda_3 {\cal M}_3 \Lambda_3^T,\qquad
{\cal A}_1 \rightarrow \Lambda_3 {\cal A}_1,\qquad
A_1 \rightarrow \Lambda_3 A_1,\nn\\
&&A_2 \rightarrow (\Lambda_3^{- 1})^T A_2,\qquad G_3 
\rightarrow  (\Lambda_3^{- 1})^T G_3,
\eqaend
where $\Lambda_3$ is a global  $SL(3,R)$ matrix. 

As we have demonstrated above, the $SL(3, R)$ group 
is indeed a global symmetry of the
theory which is realized at the level of Lagrangian. The corresponding
discrete subgroup $SL(3, Z)$ must contain all the non-perturbative U-duality
symmetries since it transforms the ten dimensional type IIB dilaton $\phi$ and 
in particular, it contains transformations which reverse the sign of the 
dilaton, while the discrete subgroup $SL(2, Z)$ of the $SL(2, R)$ does not 
transform the dilaton at all (see the following discussion). 

The $SL(2, R)$ can be 
fully realized only at the level of equations of motion since it rotates the 
equation of motion and the Bianchi identity for the 4-form field strength 
$F_4$. The above Lagrangian also implies that the $SL(2, R)$ 
transforms only the scalars ($\sigma, \rho$) while leaves the rest of the 
scalars inert (This will not be true if the type IIA basis is used instead). 
Therefore, the discrete $SL(2, Z)$ acts like the usual S-duality 
$SL(2,Z)$  as an electric/magnetic duality symmetry but unlike the S-duality 
$SL(2,Z)$ it is merely a T-duality symmetry.
Even though the $SL(2, R)$ has long been conjectured to be a symmetry of 
the $D = 8$ supergravity by Cremmer and Julia and its discrete subgroup 
$SL(2, Z)$ is 
believed to be one of $SL(2,Z)$ factors in the T-duality symmetry 
$O(2,2;Z)\equiv [SL(2,Z)\times SL(2,Z)]/Z_2$ for the $T^2$-compactified 
type II string theory, to our knowledge this $SL(2, R)$ invariance has
not been demonstrated explicitly at the level of equations of motion of the
supergravity. We postpone to give a  demonstration of this symmetry 
explicitly elsewhere [18]. When 
the 1-form potentials $A_1$ and ${\cal A}_1$ are both set to zero, showing 
the $SL(2, R)$ symmetry is not different from that of the classical 
S-duality $SL(2,R)$ in $D = 4$ [5,19]. 
Actually, Izquierdo et al [12] just employed both the $SL(2, R)$ and the 
$SL(2, Z)$ symmetries to construct the dyonic membranes in the case of 
vanishing $A_1$ and ${\cal A}_1$.

Given that the $SL(2,R)$ is indeed a symmetry of the $D = 8$ supergravity 
theory, can we construct the scalar coset 
matrix ${\cal M}_2$ in a similar fashion as we did 
for the matrix ${\cal M}_3$? The answer turns out to be true in general 
whenever we have 
field strengths which transform among themselves 
(without the need of introducing their duals) 
in a representation of the underlying group.
In other words, one should be able to see that  certain terms containing 
these field strengths in the Lagrangian  are invariant under the underlying 
global symmetry transformation. In the present case, we know that the 1-from
potential $A_1$ and ${\cal A}_1$ each transform as a triplet of the global 
$SL(3, R)$ while some combination of the two transforms as a doublet of the 
$SL(2, R)$. Examining the kinetic terms for ${\tilde F}_2$ and 
${\tilde {\cal F}}_2$ in the 
Lagrangian already suggests to us the similarity with what we know about the
strong-weak $SL(2,R)$ case in $D = 10$ type IIB theory [9]. If we write
${\cal M}_3 = \nu \nu^T$ with
\begineq
\nu = e^{\frac{\varphi}{2\sqrt 3}} \left(\begin{array}{ccc}
          e^{-\phi/2}&  \chi_3 e^{\phi/2}&(\chi_2 + \chi_1 \chi_3) 
e^{- \frac{\sqrt 3}{2} \varphi}\\
0&e^{ \phi/2} & \chi_1 e^{-\frac{\sqrt 3}{2} \varphi}\\
0&0&e^{- \frac{\sqrt 3}{2} \varphi}\end{array}\right),
\endeq
and introduce a 2-form  doublet 
\begineq
f_2 = \left(\begin{array}{c}
            \nu^{-1} {\cal F}_2\\
  	    \nu^{-1} F_2 \end{array}\right),
\endeq
then the kinetic terms for both ${\tilde {\cal F}}_2$ and ${\tilde F}_2$ 
in the Lagrangian can be written in the following simple compact form as,
\begineq
-\frac{1}{4} e\, f_2^T {\cal M}_2 f_2,
\endeq
where the scalar matrix ${\cal M}_2$ parametrizes the coset $SL(2, R)/SO(2)$
 and is given precisely by Eq.(21). The above compact form is invariant 
under the 
following $SL(2,R)$ transformations:
\begineq
g_{\mu\nu} \rightarrow g_{\mu\nu},\qquad {\cal M}_2 \rightarrow \Lambda_2 
{\cal M}_2 \Lambda_2^T,\qquad f_2 \rightarrow (\Lambda_2^T)^{-1} f_2,
\endeq
where $\Lambda_2$ is a global $SL(2, R)$ element.  

In summary, the bosonic action of the $D$-dimensional supergravity  
in $10 > D \ge 4$, obtained from either $D = 11$ supergravity or $D = 10$ type 
IIB supergravity by the dimensional reduction on torus, can be written in a 
manifest Cremmer-Julia symmetry invariant form if this symmetry is realized at 
the level of action, through redefining various fields algebraically and 
performing certain necessary dualizations for field strengths. This process also
determines the corresponding scalar coset matrix as demonstrated in the above 
for ${\cal M}_3$ in $D = 8$. If the underlying Cremmer-Julia symmetry
cannot be fully realized at the level of action, we should seek certain 
field strengths which transform among themselves (without need to 
introduce their duals) in a certain representation of the symmetry. 
Then the kinectic terms for 
these field strengths can be written in an  invariant form of 
the symmetry. This can also be used to determine the scalar coset matrix 
as we did above for
the $SL(2, R)$ case. These scalar coset matrices are important for us to 
construct U-duality multiplets in $10 > D \ge 4$ in section 6. 
In the following two sections, we will present two explicit examples of
the construction of U-duality p-brane multiplets in $D = 8$. 
First, we will show
how to construct $SL(3,Z)$ strings (also the magnetic dual 3-branes) and
then we show how to obtain $SL(3,Z) \times SL(2,Z)$ 0-branes (and the dual
4-branes) in this theory. These examples demonstrate the general features
of U-duality p-branes of various supergravity theories in diverse dimensions.

\section{$SL(3,Z)$ Strings and 3-Branes}

In this case, we need to keep the metric $g_{\mu\nu}$, 
the scalars parametrizing the scalar coset matrix ${\cal M}_3$ and the three 
2-form gauge potentials $A_2^{(i)}$ in the Lagrangian (16) or (20). The rest of 
fields in the
Lagrangian can be consistently set to  zero.
The corresponding action can then be written as follows:
\eqabegin
S_8 &=& \int\,d^8x \left[{\sqrt {-g}}\left(R + \frac{1}{4} {\rm tr}\,
\nabla_\mu {\cal M}_3 \nabla^\mu {\cal M}_3^{-1} - \frac{1}{12} {\tilde F}_3^T
{\cal M}_3 {\tilde F}_3\right)\right.\nn\\
& & \qquad\qquad\left. -\frac{1}{2.6^3}\epsilon^{\mu_1\ldots \mu_8}
{\tilde F}_{\mu_1\mu_2\mu_3}^{(i)} {\tilde F}_{\mu_4\mu_5\mu_6}^{(j)} 
A_{\mu_7\mu_8}^{(k)} \epsilon_{ijk}\right]
\eqaend
where all the notations are explained in the previous section.
As mentioned in the previous section, the $SL(2, R)$ factor 
in the Cremmer-Julia 
$SL(3, R)\times SL(2, R)$ symmetry is merely a classical T-duality 
symmetry while the 
$SL(3, R)$ contains all the classical non-perturbative U-duality symmetry.
Actually, the $SL(3, R)$ contains a strong-weak $SL(2, R)$ 
and a T-duality $SL(2, R)$
as its subgroups. This $SL(2, R)$ along with the T-duality
$SL(2, R)$ symmetry just mentioned forms the complete classical 
T-duality group 
$SL(2, R) \times SL(2, R) 
\simeq  SO(2,2)$ of the eight dimensional theory. The strong-weak $SL(2, R)$ is
actually being  inherited from the ten dimensional
type IIB theory. One way to understand the nature of the two $SL(2, R)$
 subgroups 
is to examine the scalar coset matrix ${\cal M}_3$ in (19). If we set 
$\varphi = 
\chi_1 = \chi_2 = 0$, then ${\cal M}_3$ is equivalent to the ten 
dimensional type IIB  scalar matrix parametrizing the strong-weak coset 
$SL(2, R)/SO(2)$ since $\chi_3$ is the RR scalar $\chi$ in the type IIB theory.
In order to see the T-duality $SL(2, R)$ subgroup, we cannot simply set the 
dilaton to zero. What we should do instead is to set the shifted dilaton 
to zero since it is well
known that the dilaton is shifted under T-duality. In the present context,
the shifted dilaton is ${\tilde \phi} = \phi - {\varphi}/{\sqrt 3}$ which 
is proportional to the eight dimensional dilaton as we will see.
If we set ${\tilde \phi} = \chi_2 = \chi_3 =0$, then the ${\cal M}_3$ is
equivalent to a scalar matrix which does not involve the new dilaton 
${\tilde \phi}$ and at the same time parametrizes a $SL(2, R)/SO(2)$ coset.
Therefore this $SL(2,R)$ must correspond to a T-duality $SL(2, R)$.  
If we set fields $A_2^{(2)} = A_2^{(3)} = 0$ and 
$\chi_1 = \chi_2 = \chi_3 = 0$, then the action (27) is reduced to
\begineq
S_8 = \int  d^8x {\sqrt {- g}}\left[R - \frac{1}{2} (\partial \Phi)^2 - 
\frac{1}{12} e^{-\frac{2}{\sqrt 3} \Phi} ({\tilde F}_3^{(1)})^2 
- \frac{1}{2} (\partial \Psi)^2\right],
\endeq
where we have made the field redefinitions: 
\eqabegin
\Phi &=& \frac{\sqrt 3}{2} \phi - \frac{1}{2} \varphi,\nn\\
\Psi &=& \frac{1}{2} \phi + \frac{\sqrt 3}{2}\varphi,\nn\\
\eqaend
with $\Phi$ the eight dimensional dilaton. It is easy to see that the NSNS
string solution considered in section 2 continues to be a NSNS string 
solution of the above action with $\Psi = 0$.

To construct the $SL(3, Z)$ strings (or U-duality p-branes in general), 
we always start with zero asymptotic values for the scalars, i.e., here 
${\cal M}_{30} = I$ with $I$ the unit matrix, and a pure NSNS string 
(or a pure NSNS p-brane). Here ${\cal M}_3$ is denoted as 
${\cal M}_{30}$ when the scalars take their asymptotic values, i.e., the 
subscript `0' denotes the asymptotic value. Depending on
the charge carried by the NSNS string to be a quantized unit charge or just
an arbitrary classical one, there exist two methods which can be used to 
construct the $SL(3,Z)$ strings. In the former case,  a compensating factor 
needs to be 
introduced to the initial unit charge by hand such that the transformed 
charge triplet obtained by a partially given classical 
$SL(3, R)$ transformation acting on the
initial charge triplet, can remain to be quantized. In the latter case,  
an initial charge triplet with the arbitrary classical NSNS charge 
as its only non-vanishing component is transformed by the same $SL(3,R)$ 
transformation to a general charge triplet. Then we impose the charge 
quantization on the transformed charge triplet due to the existence 
of 3-branes, the magnetic duals of strings. 
The two methods produce the same general $SL(3, Z)$ string solution
but they have different implications. For the former method, we sandwich a 
classical $SL(3, R)$ transformation between quantum mechanically allowable 
intial and final string configurations. As a consequence, 
the mass of the final string configuration is different from that of the 
 the initial configuration by the compensating factor introduced by hand
 while the $SL(3, R)$
transformation preserves the mass. This bizarre phenomenon is entirely due to
the unnatural use of the method which requires to introduce the compensating
factor by hand. We do not have this problem with the second method. Therefore,
we will employ it to construct the $SL(3,Z)$ strings in the following.

We first seek a most general $SL(3, R)$ transformation 
$\Lambda_{30}$ such that it maps the zero 
asymptotic values of the scalars to arbitrary given ones, i.e., mapping
${\cal M}_{30} = I$ to ${\cal M}_{30} = \Lambda_{30}  I \Lambda_{30}^T  
= \Lambda_{30} \Lambda_{30}^T$. Note that we can write  in general
$\Lambda_{30} = \nu_{30} R $ 
with $\nu_{30}$ a $3\times 3$ matrix in the coset $SL(3, R)/SO(3)$ and $R$ a 
$3\times 3$ $SO(3)$ matrix. Using the facts that in general 
${\cal M}_3 = \nu \nu^T$ and $R R^T = R^T R = I$, we must have, from the 
above and from Eq.(23)  with scalars taking their asymptotic  values,
\begineq
\nu_{30} = e^{\frac{\varphi_0}{2\sqrt 3} } \left(\begin{array}{ccc}
  e^{-\phi_0/2}& \chi_{30} e^{\phi_0 /2}& (\chi_{20} + \chi_{10} \chi_{30}) 
e^{-{\sqrt 3}\varphi_0 /2}\\
0&e^{\phi_0 /2} &  \chi_{10} e^{-{\sqrt 3}\varphi_0 /2}\\
0&0&e^{-{\sqrt 3}\varphi_0 /2}\end{array}\right).
\endeq
The explicit form of the $SO(3)$ matrix $R$ is not needed in what 
follows but we here write it in any case in terms of the three Euler angles 
$(\alpha, \beta, \gamma)$:
\begineq
R  =\left(\begin{array}{ccc}
\cos \alpha \cos \beta & -\sin\alpha\cos\gamma - \cos\alpha\sin\beta\sin\gamma
& \sin\alpha\sin\gamma - \cos\alpha\sin\beta\cos\gamma\\
\sin\alpha\cos\beta & \cos\alpha\cos\gamma - \sin\alpha\sin\beta\sin\gamma
&-\cos\alpha\sin\gamma - \sin\alpha\sin\beta\cos\gamma\\
\sin\beta & \cos\beta\sin\gamma &\cos\beta\cos\gamma \end{array}\right).
\endeq 

As discussed in section 2, the NSNS string configuration, carrying 
an arbitrary classical charge $Q_{(q_1,q_2,q_3)} = 
\Delta_{(q_1, q_2, q_3)}^{1/2} Q_0$ with $\Delta_{(q_1, q_2, q_3)}$ an as yet 
undetermined dimensionless factor and $Q_0$ the charge unit which may be 
taken as the quantized unit charge, is associated with
the non-vanishing NSNS gauge potential $A_2^{(1)}$. The general string 
configuration which we are going to construct requires all three 2-form 
gauge potentials $A_2^{(1)}, A_2^{(2)}, A_2^{(3)}$  to be non-zero. 
Associated with this configuration is a Noether (or electric-like) charge 
triplet 
\begineq
{\cal Q} \equiv \left(\begin{array}{c} Q^{(1)} \\ Q^{(2)} \\
Q^{(3)}\end{array}\right),
\endeq
where 
\begineq
Q^{(i)} = \int_{S^5} \left(({\cal M}_3)_{ij}\ast{\tilde F}_3^{(j)} + 
\frac{1}{2} \epsilon_{ijk} A_2^{(j)}\wedge {\tilde F}_3^{(k)}\right),
\endeq
with $S^5$ the asymptotic 5-sphere. It follows that the charge triplet  
should transform as ${\cal Q} \rightarrow \Lambda_3 {\cal Q}$. Therefore, with 
the $SL(3, R)$ transformation 
$\Lambda_{30} = \nu_{30} R$ 
acting on the initial NSNS charge, we have the following transformed charges:
\eqabegin
Q^{(1)} &=& (\Lambda_{30})_{11} \Delta_{(q_1, q_2, q_3)}^{1/2} Q_0\nn\\
&=&\Bigg[e^{-\phi_{0}/2 + \varphi_0 /{2\sqrt 3}}\, R_{11} + 
\chi_{30} e^{\phi_{0}/2 + \varphi_0 /{2\sqrt 3}}\, R_{21}\nn\\
&\,& \qquad + \left(\chi_{20} + \chi_{10}\chi_{30}\right) 
e^{-{\sqrt 3}\varphi_0 /2} \,R_{31} \Bigg] 
\Delta_{(q_1, q_2, q_3)}^{1/2} \,Q_0,\nn\\
Q^{(2)} &=& (\Lambda_{30})_{21} \Delta_{(q_1, q_2, q_3)}^{1/2} Q_0\nn\\
&=&\left[e^{\phi_{0} /2 + \varphi_0 /{2\sqrt 3}}\, R_{21}  +
\chi_{10} e^{-\varphi_0 /{\sqrt 3}}\, R_{31}\right] 
\Delta_{(q_1, q_2, q_3)}^{1/2} \,Q_0\nn\\
Q^{(3)} &=& (\Lambda_{30})_{31} \Delta_{(q_1, q_2, q_3)}^{1/2} Q_0\nn\\
&=& e^{-\varphi_0 /{\sqrt 3}}\, R_{31} \Delta_{(q_1, q_2, q_3)}^{1/2} \,Q_0
\eqaend
Given the vacuum moduli and the three charges $Q^{(i)}$ ($i = 1, 2, 3$), 
we have three $SO(3)$ group 
parameters and the additional $\Delta_{(q_1,q_2,q_3)}$ to be fixed.
However, we have only three equations in (34). This is in contrary 
to the case 
for the $SL(2, Z)$ strings [9] or fivebranes [11] of $D = 10$ 
type IIB theory 
where under the similar conditions the $SL(2, R)$ parameters are 
completely fixed. Surprisingly,  we find that
the most important factor $\Delta_{(q_1,q_2,q_3)}$ can 
nevertheless be completely determined as will be demonstrated below. Then
 it can be seen from (34) that we can only determine two of the 
three $SO(3)$ group parameters. This seems to 
imply that our general string solution will contain an arbitrary parameter. 
But again to our surprise we find that the general solution has nothing to 
do with this
arbitrary group parameter and all the relevant physical quantities can be
uniquely determined as we will show below. 

Solving the $SO(3)$ matrix elements $R_{11}, R_{21}, R_{31}$ from (34), 
we have
\eqabegin
R_{11} &=& e^{\phi_0 /2 - \varphi_0 /{2\sqrt 3}} \,
\Delta_{(q_1,q_2,q_3)}^{- 1/2}\, \frac{Q^{(1)} - \chi_{30} Q^{(2)} - 
\chi_{20} Q^{(3)}}{Q_0},\nn\\
R_{21} &=& e^{-\phi_0 /2 - \varphi_0 /{2\sqrt 3}}\, 
\Delta_{(q_1,q_2,q_3)}^{- 1/2}\, \frac{Q^{(2)} - \chi_{10} Q^{(3)}}{Q_0},\nn\\
R_{31} &=& e^{\varphi_0 /{\sqrt 3}}\, \Delta_{(q_1,q_2,q_3)}^{- 1/2} 
\,\frac{Q^{(3)}}{Q_0}.
\eqaend
Using the orthogonal relation $R_{ki} R_{kj} = \delta_{ij}$ for $i = j = 1$,
i.e., $R_{11}^2 + R_{21}^2 + R_{31}^2 = 1$, we can fix the 
$\Delta_{(q_1,q_2,q_3)}$ as
\eqabegin
\Delta_{(q_1,q_2,q_3)} &=& e^{2\varphi_0 /{\sqrt 3}} 
\left(\frac{Q^{(3)}}{Q_0}\right)^2 + e^{-\phi_0 - \varphi_0 /{\sqrt 3}}
\left(\frac{Q^{(2)} - \chi_{10} Q^{(3)}}{Q_0}\right)^2 \nn\\
&\,&\qquad + e^{\phi_0 - \varphi_0 /{\sqrt 3}} 
\left(\frac{Q^{(1)} - \chi_{30} Q^{(2)}
-\chi_{20} Q^{(3)}}{Q_0}\right)^2,\nn\\
&=& \left(Q^{(1)}/Q_0, Q^{(2)}/Q_0,Q^{(3)}/Q_0\right) 
{\cal M}_{30}^{-1} \left(\begin{array}{c}
Q^{(1)}/Q_0\\ Q^{(2)}/Q_0\\ Q^{(3)}/Q_0\end{array}
\right),
\eqaend
where
\begineq
{\cal M}_{30}^{-1} = e^{-\frac{\varphi_0}{\sqrt 3}}
\left(\begin{array}{ccc} e^{\phi_{0}} &
-\chi_{30} e^{\phi_{0}} & -\chi_{20} e^{\phi_{0}}\\
-\chi_{30} e^{\phi_{0}} & \chi_{30}^2 e^{\phi_{0}} + e^{- \phi_{0}} &
 -\chi_{10} e^{- \phi_{0}} + \chi_{20}\chi_{30} e^{\phi_{0}}\\
-\chi_{20} e^{\phi_{0}} & \chi_{20}\chi_{30} e^{\phi_{0}} - \chi_{10}
e^{- \phi_{0}} & \chi_{20}^2 e^{\phi_{0}} + \chi_{10}^2 
e^{- \phi_{0}} + e^{{\sqrt 3} \varphi_0}\end{array}\right).
\endeq
{}From (36), it is clear that $\Delta_{(q_1, q_2, q_3)}$ is $SL(3, R)$
invariant.  

By now we have constructed a most general $D = 8$ string 
configuration carrying classical charges given by the charge triplet. 
The central
charge (therefore the ADM mass per unit length as well as 
the tension measured in Einstein
metric) associated with this string is 
$Q_{(q_1,q_2,q_3)} = \Delta_{(q_1,q_2,q_3)}^{1/2} Q_0$ with 
$\Delta_{(q_1,q_2,q_3)}$ as given in Eq.(36). The 
metric continues to be given by the one in Eq.(3) but now with $Q = 
Q_{(q_1,q_2,q_3)}$.
The three 3-form field strengths are now given by the triplet 
\eqabegin
\left(\begin{array}{c}
{\tilde F}_3^{(1)}\\{\tilde F}_3^{(2)}\\{\tilde F}_3^{(3)}\end{array}\right)
&=& (\Lambda_{30}^T)^{-1} \left(\begin{array}{c}
\Delta_{(q_1,q_2,q_3)}^{1/2}\, Q_0\, A^{-1/{\sqrt 3}} (\rho)\, *\epsilon_5\\
0\\0\end{array}\right),\nn\\
&=& {\cal M}_{30}^{- 1} \left(\begin{array}{c} Q^{(1)}\\Q^{(2)}\\Q^{(3)}
\end{array}\right) A^{-\frac{1}{\sqrt 3}} (\rho) \ast\epsilon_5,
\eqaend
where we have used ${\cal M}_{30}^{- 1} = ({\Lambda^T}_{30})^{-1} 
\Lambda_{30}^{-1}$,
\begineq
{\cal Q} = \Lambda_{30} \left(\begin{array}{c}
\Delta_{(q_1, q_2, q_3)}^{1/2} Q_0\\
0\\
0\end{array}\right),
\endeq
 and $A(\rho)$ is given in Eq.(3). So far all the 
above quantities are independent of the undetermined arbitrary $SO(3)$ group 
parameter. 
Our last step to complete 
the construction of the general classical string solution is to determine all
the scalars appearing in ${\cal M}_3$ as given by Eq.(19). This can be 
achieved
by the following matrix equation:
\eqabegin
{\cal M}_3 &=& A^{-\frac{1}{2\sqrt 3}} (\rho) \Lambda_{30} 
\left(\begin{array}{ccc}
     A^{\frac{\sqrt 3}{2}} (\rho)&0&0\\
     0&1&0\\
     0&0&1\end{array}\right) \Lambda_{30}^T,\\
&=&A^{-\frac{1}{2\sqrt 3}} (\rho) \nu_{30} \left(\begin{array}{ccc}
R_{11}^2 B(\rho) + 1&R_{11} R_{21} B(\rho) &R_{11} R_{31} B(\rho)\\
R_{21} R_{11} B(\rho)& 
R_{21}^2 B(\rho) + 1&
R_{21} R_{31} B(\rho)\\
R_{31} R_{11} B(\rho)& 
R_{31} R_{21} B(\rho)&
R_{31}^2 B(\rho) + 1\end{array}\right) \nu_{30}^T,
\eqaend
where $B(\rho) = A^{{\sqrt 3}/2} (\rho) - 1$.
Let us examine a few things here. As $\rho \rightarrow \infty$, 
$A (\rho) \rightarrow 1$ and $B (\rho) \rightarrow 0$. 
So, from (41), we first have 
${\cal M}_3 \rightarrow \nu_{30} \nu_{30}^T = {\cal M}_{30}$ as expected.
Second, the right side of (41) is completely fixed, 
independent of the undetermined $SO(3)$ group parameter, since 
$R_{11}, R_{21}, R_{31}$ are completely fixed by Eq.(35). 
Furthermore, the scalars 
appearing in ${\cal M}_3$ can be determined  without even a sign 
ambiguity. Look 
at the structure of ${\cal M}_3$ matrix in (19). From the above equation, 
we can
simply read out $\varphi$ first. We can then fix $\chi_1$. Then $\phi$ and 
$\chi_2 + \chi_1 \chi_3$. With all these known, we can then determine 
$\chi_3$. Applying this $\chi_3$ and the known $\chi_1$ to the known 
$\chi_2 + \chi_1 \chi_3$, we finally fix $\chi_2$. It follows that all scalars
for this solution are independent of the undetermined $SO(3)$ parameter. We 
therefore confirm our claim that the $SL(3,Z)$ multiplet of string solutions
can be obtained without any arbitrariness eventhough one of the $SO(3)$
parameter remains undetermined. We will not present the explicit 
expressions for each of the scalars here. 

Our general classical string 
solution also preserves half of the spacetime supersymmetry as the original 
pure NSNS string since the global $SL(3,R)$ transformation 
commutes with the
supersymmetry transformation. Therefore, our general string solution 
continues to be BPS
which implies that the ADM mass per unit length, the central charge 
$Q_{(q_1,q_2,q_3)}$ and the string tension measured in Einstein metric are all
the same in proper units.
  
So far we have only constructed the most general classical string solution in
the sense that the three charges $Q^{(1)}, Q^{(2)}, Q^{(3)}$ can be arbitrary.
Due to the presence of the magnetic duals of strings, i.e., the 3-branes, each 
of the three charges must be quantized [14]
separately in terms of the unit charge $Q_0$.
For example, the magnetic-like charges $P^{(1)} \neq 0, 
P^{(2)} = 0, P^{(3)} = 0$ carried by a 3-brane must imply that $Q^{(1)}$ is 
quantized in terms of the unit charge $Q_0$. So the charge triplet for
a general quantum-mechanically allowable string solution is
\begineq
{\cal Q} = \left(\begin{array}{c} q_1\\q_2\\q_3\end{array}\right) Q_0,
\endeq
where $q_1, q_2, q_3$ are three integers. In terms of the unit charge $Q_0$, 
the charge triplet should remain to be an integral triplet under 
quantum-mechanically allowable transformation. This necessarily breaks the 
continuous $SL(3, R)$ symmetry to a discrete $SL(3, Z)$ whose elements take only
integral values. 
 
The most general quantum-mechanically allowable string configuration can be 
obtained simply by imposing $Q^{(1)} = q_1 Q_0, Q^{(2)} = q_2 Q_0, Q^{(3)} =
q_3 Q_0$ in the above classical string configuration. For example, 
\begineq
\Delta_{(q_1,q_2,q_3)} = \left(q_1, q_2, q_3\right) 
{\cal M}_{30}^{-1} \left(\begin{array}{c}
q_1\\ q_2\\ q_3\end{array}
\right), 
\endeq
which is now $SL(3, Z)$ invariant. Therefore, the ADM mass per unit length 
$M_{(q_1,q_2,q_3)}$, the central charge $Q_{(q_1,q_2,q_3)}$ and the string 
tension $T_{(q_1,q_2,q_3)}$ measured in Einstein metric are all $SL(3, Z)$
invariant. In proper units, we can set all three equal in which case we can
take $Q_0$ as the fundamental string tension $T$. Then for a 
$(q_1,q_2,q_3)$-string, we have
\eqabegin
M_{(q_1,q_2,q_3)} &=& Q_{(q_1,q_2,q_3)} = T_{(q_1, q_2, q_3)} =
 \Delta _{(q_1, q_2, q_3)}^{1/2} T\nn\\
&=& \sqrt{e^{\frac{2\varphi_0}{\sqrt 3}} 
q_3^2 + e^{-\phi_0 - \frac{\varphi_0}{\sqrt 3}}
\left(q_2 - \chi_{10} q_3\right)^2 +
e^{\phi_0 - \frac{\varphi_0}{\sqrt 3}} \left(q_1 - \chi_{30} q_2
-\chi_{20} q_3\right)^2 }\,T
\eqaend
The $(q_1, q_2, q_3)$-string tension measured in string metric is
\eqabegin
T_{(q_1, q_2, q_3)} &=& e^{- \Phi_0 /{\sqrt 3}} \Delta_{(q_1,q_2,q_3)}^{1/2} T
\nn\\
&=& \sqrt{e^{-\phi_0 + {\sqrt 3} \varphi_0} q_3^2 + e^{ - 2 \phi_0} 
(q_2 - \chi_{10} q_3)^2 + (q_1 - \chi_{30} q_2 - \chi_{20} q_3)^2}\,T
\eqaend
where $\Phi_0 = {\sqrt 3} \phi_0 /2 - \varphi_0 /2$. 

Let us make a few comments about the above 
tension formula. For simplicity, we set $\chi_{10} = \chi_{20} = 
\chi_{30} = 0$. We note that the tension for $(1,0,0)$-string 
is proportional to 1 which is
expected since this is a NSNS string. The tension for $(0, 1, 0)$-string is 
proportional to $ e^{ - \phi_0} = 1/ g_s$, i.e., inversely to the $ D = 10$
type IIB string coupling constant. This is also expected since this string 
is a D-string [20] in $D = 10$ and this tension relation can be easily 
verified through simple 
dimensional reduction of the $D = 10$ D-string $\sigma$-model action to 
$D = 8$. The tension for
the $(0, 0, 1)$-string is, however, proportional to 
$e^{ - \phi_0/2 + {\sqrt 3}\varphi_0 /2} = 
e^{- (\Phi_0 - \varphi_0) /{\sqrt 3}}$, a strange
behavior. We would naively expect that this is also a D-string, i.e., the 
tension should be inversely proportional to either the $D = 10$ string coupling
constant or $D = 8$ string coupling constant which is $e^{\sqrt{3} \Phi_0}$ [6].
It turns out that this string is a $D = 10$ type IIB D-threebrane with two
of its spatial dimensions wrapped on the 
two compactified dimensions when we go from 10
to 8. 

Let us see how this tension can be obtained from the $D = 10$ D-threebrane 
$\sigma$-model action.  In $D = 10$
the action for the D-threebrane to the lowest order (ignoring the world volume
vector fields) can be represented in 
string metric as
\begineq
 S_3 \sim \int d^4 \xi\, e^{ - \phi} \sqrt{-\gamma} \gamma^{ij} \partial_i X^M 
\partial_j X^N g_{MN},
\endeq
where the factor $e^{ - \phi}$ indicates that the threebrane tension is 
inversely proportional to the string coupling constant, i.e., $\sim 1/g_s$, 
 and the so-called worldvolume induced metric $\gamma_{ij}
= \partial_i X^M \partial_j X^N g_{MN}$ with $i = 0, 1, 2, 3$. 
In order to obtain the 
$(0,0,1)$-string in $D = 8$ from the threebrane in $D = 10$, we have to adopt 
the so-called double dimensional reduction procedure [21], i.e., 
identifying two
worldvolume spatial dimensions with the two compactified spatial dimensions
of spacetime, i.e., $\xi^3 = z^9, \xi^2 = z^8$ with $z^8, z^9$ the two 
compactified space-like dimensions of spacetime. 
When we compactify the $D = 10$
type IIB supergravity theory to $D = 8$, the Einstein metric is
\begineq
d s_{10}^2 = e^{ - \varphi /2{\sqrt 3}} d s_8^2 + e^{{\sqrt 3}\varphi /2} 
\left[ (d z^8)^2 + (d z^9)^2\right],
\endeq
where $d s_8^2$ is the eight dimensional Einstein metric and the Kaluza-Klein
vectors are ignored here.
The ten dimensional string metric (also the eight dimensional string metric)
is
\eqabegin
d s^2_{10} ({\rm string\, metric}) &=& e^{\phi/2} \,d s^2_{10},\nn\\
&=& e^{\Phi/{\sqrt 3}} d s^2_8 + e^{\phi/2 + {\sqrt 3}\varphi/2}  
 \left[ (d z^8)^2 + (d z^9)^2\right],
\eqaend
where $\Phi$ is the eight dimensional dilaton. With these 
metric relations and 
assuming that all the fields are independent of $z^8$ and $z^9$,  
we have $\gamma_{22} = \gamma_{33} =  e^{\phi/2 + {\sqrt 3}\varphi/2}$.
Now the D-threebrane action in $D = 10$ goes to the $(0,0,1)$-string action 
in $D = 8$ as
\begineq
S_3 \rightarrow S_1 \sim \int d^2 \xi e^{- \phi} \gamma_{22} \sqrt{-\gamma}
 \gamma^{ij} \partial_i X^\mu  \partial_j X^\nu g_{\mu \nu},
\endeq
where $i = 0, 1$. From the above, we have the $(0, 0, 1)$-string tension 
proportional to 
$e^{- \phi_0} (\gamma_{22})_0 = e^{- \phi_0 /2 + {\sqrt 3} \varphi_0/2}$, 
which is exactly the same as that given by our tension formula (45).

As discussed earlier, the global $SL(3, R)$ contains a strong-weak $SL(2,R)$ 
subgroup (corresponding to $\varphi = \chi_1 = \chi_2 = 0$) 
and a T-dual $SL(2,R)$ subgroup (corresponding to  
${\tilde \phi} = \chi_2 = \chi_3 = 0$). Similarly, we expect that the quantum
$SL(3,Z)$ contains a strong-weak $SL(2,Z)$ subgroup and a T-dual $SL(2,Z)$
subgroup. Evidence for this can be provided by the tension formula (45).
When $\varphi = \chi_1 = \chi_2 = 0$, we recover the tension formula for
the type IIB $SL(2,Z)$ $(q_1, q_2)$-string as discussed in ref.[9]. Also
for ${\tilde \phi} = \chi_2 = \chi_3 = 0$, 
we have the formula for the T-dual $SL(2, Z)$ $(q_2, q_3)$-strings. For the 
$(q_2,q_3)$-string, the tension for $(q_2,0)$-string is inversely proportional
to the tension for $(0, q_3)$-string. This inverse relation is actually 
$1/R \rightarrow R $, a typical T-duality relation, with 
$R = e^{\varphi_0 /{\sqrt 3}}$ the compactification radius measured in string 
metric. In other words, the $(q_2, 0)$-string carries momentum modes while
$(0,q_3)$-string carries winding modes with respect to the compactifications.   
 
The 3-form field strength triplet is now given as,
\begineq
{\tilde F}_3 = {\cal M}_{30}^{- 1} \left(\begin{array}{c} q_1\\q_2\\q_3
\end{array}\right) Q_0 A^{-\frac{1}{\sqrt 3}} (\rho) *\epsilon_5,
 \endeq
As mentioned earlier, the metric in Eq.(3) retains the same form but now 
with $Q = Q_{(q_1,q_2,q_3)}$:
\begineq
ds^2 = \left(1 + \frac{Q_{(q_1, q_2, q_3)}}{4 \rho^4}\right)^{-2/3}
\left[- dt^2 + (dx^1)^2\right] + \left(1 + \frac{Q_{(q_1, q_2, q_3)}}
{4\rho^4}\right)^{1/3} \left[d\rho^2 + \rho^2 d\Omega_5^2\right]
\endeq

The above $(q_1,q_2,q_3)$-string configuration encodes all the information 
about the $SL(3, Z)$ multiplets of the $D = 8$ strings. 
Note that for given asymptotic values
of the scalars, i.e., for a given vacuum, each of the infinitely many 
integral triplets 
$(q_1,q_2,q_3)$ gives a different value for the $\Delta_{(q_1,q_2,q_3)}$ which
cannot be related to each other by a $SL(3, Z)$ transformation since it is 
invariant by such a transformation. Further, this $\Delta_{(q_1,q_2,q_3)}$ 
measures the mass per unit length, the central charge
and the tension. Therefore, we can use this factor to label different 
$SL(3, Z)$
multiplets. Within each such multiplet, we have a collection of infinitely many 
discrete vacua and a collection of infinitely many integral charge triplets. 
Each of such vacua and its corresponding integral charge triplet are obtained
from the given initial vacuum ${\cal M}_{30}$ and the 
given initial charge triplet $(q_1, q_2, q_3)$ by a particular 
$SL(3, Z)$ transformation. 
Picking a special vaccum in such a multiplet will break the $SL(3, Z)$ 
spontaneously. In other words, all the string configurations in such a 
multiplet are physically equivalent. The physically inequivalent string 
configurations are those with different $\Delta_{(q_1,q_2,q_3)}$ values which
correspond to different integral triplet $(q_1,q_2,q_3)$ for a fixed 
${\cal M}_{30}$, i.e., a fixed vaccum.  

Finally, we would like to discuss the stability of a general
$(q_1,\,q_2,\,q_3)$ string (Discussion of the stability for Type IIB $SL(2, Z)$
strings is given in [9,22]). We have noted that the tension of such
a string is given by $T_{(q_1, q_2, q_3)} = \Delta_{(q_1, q_2, q_3)}^
{1/2} T$ and so, it can be easily checked that the tensions satisfy
the following triangle inequality relation, irrespective of the vacuum moduli,
\begineq
T_{(q_1, q_2, q_3)} + T_{(p_1, p_2, p_3)} \geq T_{(q_1+p_1, q_2+p_2,
q_3+p_3)}
\endeq
where the equality holds if and only if $p_1 q_2 = 
p_2 q_1$, $p_2 q_3 = p_3 q_2$
and $p_1 q_3 = p_3 q_1$, i.e., when, $p_1 = n q_1$, $p_2 = n q_2$,
$p_3 = n q_3$, with $n$ being an integer. So, when any two of the three
integers $q_1, q_2, q_3$ are
relatively prime to each other the string would be stable as the 
$(q_1,\,q_2,\,q_3)$-string in that case will be prevented from decaying
by the inequality relation (52) called as the `tension gap' equation. The
same conclusion can be drawn when the central charge triangle inequality 
relation (now called `charge gap' equation) and the charge conservation are 
employed. We therefore conclude that  all physically inequivalent 
stable $(q_1,q_2,q_3)$-string configurations are those corresponding to  all 
possible integral triplets $(q_1, q_2, q_3)$ with any two of the three 
integers in each triplet relatively prime, and with ${\cal M}_{30}$
belonging to the fundamental region of $SL(3,Z)$, i.e., 
$SL(3,Z) \backslash SL(3,R)/SO(3)$. 

The magnetic dual of a string in $D = 8$ is a 3-brane. The $SL(3, Z)$ family of 
3-branes can be constructed following the same steps as we did in ref.[11] for 
the $SL(2, Z)$ fivebranes, the magnetic duals of strings in 
$D = 10$ type IIB theory. We will not repeat these steps here 
but merely present the 
$SL(3, Z)$ $(p_1, p_2, p_3)$-threebrane configuration associated with a 
magnetic-like integral charge triplet denoted as $p$. The $\Delta$-factor
in this case is, 
\begineq
\Delta_{(p_1, p_2, p_3)} = \left(p_1, p_2, p_3\right) {\cal M}_{30}
\left(\begin{array}{c}
       p_1\\
       p_2\\
       p_3\end{array}\right).
\endeq

The mass per unit 3-brane volume $M_{(p_1, p_2, p_3)}$, the central charge 
$Q_{(p_1, p_2, p_3)}$ and the tension $T_{(p_1, p_2, p_3)}$ measured in 
Einstein metric are
\eqabegin
M_{(p_1, p_2, p_3)} &=& Q_{(p_1, p_2, p_3)} = T_{(p_1, p_2, p_3)} =
\Delta_{(p_1, p_2, p_3)}^{1/2} Q_0,\nn\\
&=&\Bigg\{e^{-\phi_0 + \varphi_0/\sqrt{3}} p_1^2 + e^{\phi_0 + 
\varphi_0/\sqrt{3}} (p_2 + \chi_{30} p_1)^2 \nn\\
&\,&\,\,\,\,\, + e^{- 2\varphi_0/\sqrt{3}}\left[
p_3 + \chi_{10} p_2 + (\chi_{20} + \chi_{10} \chi_{30}) p_1\right]^2
\Bigg\}^{1/2} Q_0,
\eqaend
where $Q_0$ is the unit magnetic charge which can be taken as 
the fundamental 3-brane tension
$T$.

The 3-form field strength triplet is
\begineq
{\tilde F}_3 = \left(\begin{array}{c}
       p_1\\
       p_2\\
       p_3\end{array}\right) Q_0 \epsilon_3.
\endeq
Similarly, the scalars are determined uniquely by the following matrix equation
\eqabegin
{\cal M}_3 &=& A^{\frac{1}{2\sqrt 3}} (\rho) \Lambda_{30} 
\left(\begin{array}{ccc}
     A^{- \frac{\sqrt 3}{2}} (\rho)&0&0\\
     0&1&0\\
     0&0&1\end{array}\right) \Lambda_{30}^T,\\
&=&A^{\frac{1}{2\sqrt 3}} (\rho) \nu_{30} \left(\begin{array}{ccc}
R_{11}^2 B(\rho) + 1&R_{11} R_{21} B(\rho) &R_{11} R_{31} B(\rho)\\
R_{21} R_{11} B(\rho)& 
R_{21}^2 B(\rho) + 1&
R_{21} R_{31} B(\rho)\\
R_{31} R_{11} B(\rho)& 
R_{31} R_{21} B(\rho)&
R_{31}^2 B(\rho) + 1\end{array}\right) \nu_{30}^T,
\eqaend
where $B(\rho) = A^{-\sqrt{3}/2} (\rho) - 1$ 
with $A(\rho) = \left(1 + Q_{(p_1, p_2, p_3)}/2 \rho^2\right)^{2/\sqrt{3}}$
and 
\eqabegin
R_{11} &=& \Delta_{(p_1, p_2, p_3)}^{- 1/2} 
\,e^{-\phi_0/2 + \varphi_0 /2\sqrt{3}}\, p_1 ,\nn\\
R_{21} &=& \Delta_{(p_1, p_2, p_3)}^{- 1/2}\, e^{\phi_0/2 + \varphi_0/2\sqrt{3}}
\,\left(p_2 + \chi_{30} p_1\right),\nn\\
R_{31} &=& \Delta_{(p_1, p_2, p_3)}^{- 1/2}\, e^{-\varphi_0/\sqrt{3}} \,
\left[p_3 + \chi_{10} p_2 + \left(\chi_{20} + \chi_{10} \chi_{30}\right) p_1
\right].
\eqaend
The metric is
\begineq
ds^2 = \left(1 + \frac{Q_{(p_1, p_2, p_3)}}{2 \rho^2}\right)^{-1/3} 
\left[- (dt)^2 + (dx^i)^2\right] + 
\left(1 + \frac{Q_{(p_1, p_2, p_3)}}{2 \rho^2}\right)^{2/3} \left[
(d\rho)^2 + \rho^2\, d\Omega_3^2\right],
\endeq
with $i = 1, 2, 3$.   
In string metric, the tension for a $(p_1, p_2, p_3)$-threebrane is
\eqabegin
T_{(p_1, p_2, p_3)} &=&  \Bigg\{e^{-3\phi_0 + \sqrt{3} \varphi_0} p_1^2 + 
e^{- \phi_0 + 
\sqrt{3} \varphi_0} (p_2 + \chi_{30} p_1)^2\nn\\
&\,&\,\,\,\,\, + e^{- 2\phi_0}\left[
p_3 + \chi_{10} p_2 + (\chi_{20} + \chi_{10} \chi_{30}) p_1\right]^2
\Bigg\}^{1/2} T.
\eqaend
Using the brane $\sigma$-model action approach discussed before, we can 
understand this tension formula easily from the facts that the
$(1, 0, 0)$-threebrane is a $D = 10$ type IIB NSNS 5-brane [23] 
wrapped on the two 
compactified dimensions and $(0, 1, 0)$-threebrane a wrapped 
$D = 10$ type IIB RR 5-brane [24] while $(0, 0, 1)$-threebrane is obtained by
simple dimensional reduction of the $D = 10$ type IIB threebrane [8,25].

As for the $SL(3, Z)$ strings, similar results can also be obtained 
 on multiplets and stability for the 3-branes.     
The corresponding $SL(3, Z)$ black strings and black 3-branes can also be 
constructed similarly. 

\section{$SL(3,Z) \times SL(2,Z)$ 0-Branes and 4-Branes}

Given that $SL(3,R)\times SL(2,R)$ is the Cremmer-Julia symmetry of 
the $D = 8$ 
theory, we cannot resist to give a complete construction of all U-duality
p-branes in this theory. In this section, we will present the last two 
U-duality p-branes, namely, U-duality 0-branes and 4-branes. Let us discuss
the 0-branes first.

We will construct the $SL(3,Z)\times SL(2,Z)$ multiplets of 0-branes 
from the following 
known 0-brane configuration preserving half of the spacetime 
supersymmetry [6],
\eqabegin
ds^2 &=& - \left(1 + \frac{Q}{5 \rho^5}\right)^{- 5/6} (dt)^2 + 
\left(1 + \frac{Q}{5 \rho^5}\right)^{1/6} \left[d \rho^2 + \rho^2 d \Omega_6^2
\right],\nn\\
e^{- 2{\tilde \Phi}} &=& \left(1 + \frac{Q}{5\rho^5}\right)^{\sqrt{7/3}} =
A(\rho),\,\,\,
{\tilde {\cal F}}^{(1)}_2 = Q A^{- \sqrt{7/12}} (\rho) \ast \epsilon_6,
\eqaend
which is the solution of the following action
\begineq
S = \int d^8 x \sqrt{-g}\left[R - \frac{1}{2} (\partial {\tilde \Phi})^2 
- \frac{1}{4} \, e^{-\sqrt{7/3} {\tilde \Phi}}
({\tilde {\cal F}}_2^{(1)})^2\right].
\endeq
In order to obtain the above action from our action (16), we are forced to take
\begineq
\sigma = - \phi =  \sqrt{\frac{3}{7}}\, {\tilde \Phi},\,\, \varphi = 
\sqrt{\frac{1}{7}} \,{\tilde \Phi},
\endeq
and we also need to set  the rest of the fields  not relevant for us to 
zero. 
In other words, to have a 0-brane solution which preserves half of the
spacetime supersymmetry, we are forced to have non-vanishing $\sigma$ field.
This turns out to be the key to have a complete construction for
the $SL(3,Z)\times SL(2,Z)$ 0-brane solutions. Otherwise, 
the $SL(2,Z)$ factor would have been 
trivial. 

We could construct the $SL(3,Z)\times SL(2,Z)$ multiplets of 0-branes 
by following the same
route as we did for the $SL(3,Z)$ strings. But from the study of the $SL(3,Z)$
strings along with the examples studied previously in [9,11,12], 
we learn that 
a U-duality p-brane
configuration can be determined completely by the underlying symmetry 
properties without the need to follow the detail steps as, for example, we did 
for the $SL(3,Z)$ strings, once a particular p-brane configuration is known. 
In other words,
we can simply write down a U-duality p-brane configuration based on the 
underlying symmetry properties. We will use here the latter method to write
down the 0-brane solution. This specific example will also serve the purpose
of demonstrating the method for constructing the general U-duality p-branes 
of various supergravity theories in diverse dimensions 
which will be presented in the following section.

  Starting from the above particular 0-brane solution, we write down first 
the $SL(3,Z)$ 0-branes involving the 2-from field strengths 
${\tilde {\cal F}}^{(i)}_2$ with $i = 1, 2, 3$. 
The $\Delta_{(q_1, q_2, q_3)}$-factor for this $SL(3,Z)$ 0-brane is
\begineq
\Delta_{(q_1, q_2, q_3)} = \left(q_1, q_2, q_3\right) 
{\cal M}_{30} \left(\begin{array}{c}
                                                              q_1\\
  		                                              q_2\\
                                                              q_3\end{array}
                                                             \right),
\endeq
which is $SL(3, Z)$ invariant  as follows from (22). Here ${\cal M}_{30}$ 
is the scalar coset 
matrix given by Eq.(19) with the scalars taking their asymptotic values. 
The ADM mass and the central charge are
\begineq
M_{(q_1, q_2, q_3)} = Q_{(q_1, q_2, q_3)} = \Delta_{(q_1, q_2, q_3)}^{1/2} Q_0,
\endeq
where $Q_0$ is the unit electric charge. The 2-form field strength triplet 
is now given as,
\begineq
{\tilde {\cal F}}_2 = {\cal M}_{30} \left(\begin{array}{c}
                                                              q_1\\
  		                                              q_2\\
                                                              q_3\end{array}
                                                             \right) Q_0 
A^{-\sqrt{7/12}} (\rho) \ast \epsilon_6.
\endeq
As we did for the $SL(3,Z)$ strings, the scalars can be uniquely determined by
the following matrix equation
\eqabegin
{\cal M}_3 &=& A^{1/\sqrt{21}} (\rho)\, \Lambda_{30} \left(\begin{array}{ccc}
        A^{- \sqrt{3/7}} (\rho) & 0 &0\\
        0&1&0\\
        0&0&1\end{array}\right) \Lambda^T_{30},\nn\\
    &=& A^{1/\sqrt{21}} (\rho)\, 
 \nu_{30} \left(\begin{array}{ccc}
R_{11}^2 B(\rho) + 1&R_{11} R_{21} B(\rho) &R_{11} R_{31} B(\rho)\\
R_{21} R_{11} B(\rho)& 
R_{21}^2 B(\rho) + 1&
R_{21} R_{31} B(\rho)\\
R_{31} R_{11} B(\rho)& 
R_{31} R_{21} B(\rho)&
R_{31}^2 B(\rho) + 1\end{array}\right) \nu_{30}^T,
\eqaend
where $B (\rho) = A^{-\sqrt{3/7}} (\rho) - 1$. The same discussion as for the
$SL(3,Z)$ strings applies here. The corresponding metric for the 
$(q_1, q_2, q_3)$-particle continues to be given by the metric in (61)  but now
with $Q = Q_{(q_1, q_2, q_3)}$.

We now start to construct the $SL(3,Z)\times SL(2,Z)$ 0-brane directly from
the initial 0-brane configuration. If we denote $q$ as the integral 
electric charge 
triplet associated with the 2-from field strength triplet ${\tilde {\cal F}}_2$
and $q'$ as the integral electric charge triplet associated with the 2-form
field strength triplet ${\tilde F}_2$, we then have the 
$\Delta_{(q, q')}$-factor as
\begineq
\Delta_{(q,q')} = \left(q^T, {q'}^T\right) {\cal M}_{20}^{-1} 
\left(\begin{array}{c}
{\cal M}_{30} q\\
{\cal M}_{30} q'\end{array}\right),
\endeq
which is $SL(3,Z)\times SL(2,Z)$ invariant. Here ${\cal M}_{20}$ is 
the scalar coset matrix given by Eq.(21) with the scalars taking their 
asymptotic values.
    
The ADM mass $M_{(q, q')}$ and the central charge $Q_{(q, q')}$ are given
by
\eqabegin
M_{(q,q')} &=& Q_{(q, q')} = \Delta_{(q, q')}^{1/2} Q_0,\nn\\
  	&=& \left[e^{\sigma_0} \left(q - \rho_0 q'\right)^T {\cal M}_{30} 
 \left(q - \rho_0 q'\right) + e^{-\sigma_0} {q'}^T {\cal M}_{30} q'\right]^{1/2}
Q_0,\nn\\ 	
  	&=& \Bigg\{e^{\sigma_0 - \phi_0 + \varphi_0/{\sqrt 3}} 
\left(q_1 - \rho_0 q_1'\right)^2 + e^{\sigma_0 + \phi_0 + \varphi_0/{\sqrt 3}}	
\left[ \left(q_2 - \rho q_2'\right) + \chi_{30} \left(q_1 - \rho_0 q_1'\right)
\right]^2\nn\\
&\,&\, + e^{\sigma_0 - 2\varphi_0/{\sqrt 3}}\left[ \left(q_3 - \rho_0 q_3'\right)
 + \chi_{10}\left(q_2 - \rho_0 q_2'\right) + \left(\chi_{20} + \chi_{10} 
\chi_{30} \right) \left(q_1 -\rho_0 q_1'\right)\right]^2\nn\\
&\,& \,+ e^{-\sigma_0 - \phi_0 + \varphi_0/{\sqrt 3}} {q'}_1^2 + 
e^{-\sigma_0 + \phi_0 + \varphi_0/{\sqrt 3}} \left( q_2' + \chi_{30} q_1'
\right)^2\nn\\
&\,& \,+ e^{-\sigma_0 - 2\varphi_0/{\sqrt 3}} \left[ q_3' + \chi_{10} q_2'
+ \left(\chi_{20} + \chi_{10} \chi_{30}\right) q_1'\right]^2 \Bigg\}^{1/2} Q_0.
\eqaend
The six 2-form field strengths ${\tilde {\cal F}}_2^{(i)}$ and 
${\tilde F}_2^{(i)}$ with $i = 1, 2, 3$ are given by
\begineq
\left(\begin{array}{c}
      {\tilde {\cal F}}_2\\
      {\tilde F}_2\end{array}\right) =
{\cal M}_{20}^{-1} \left(\begin{array}{c}
                        {\cal M}_{30} q\\
                        {\cal M}_{30} q'\end{array}\right) Q_0 
 A^{-\sqrt{7/12}} (\rho) \ast \epsilon_6.
\endeq
The scalars parametrizing the coset $SL(3,R)/SO(3)$ continue to be given by
Eq.(67) but now the $SO(3)$ elements appearing in the equation take the form, 
\eqabegin
R_{11} &=& \pm \Delta_{(q,q')}^{-1/2}\, e^{-\phi_0/2 + \varphi_0/{2\sqrt 3}}
\,\left[ e^{\sigma_0} \left(q_1 -\rho_0 q_1'\right)^2 + e^{-\sigma_0} {q'}_1^2
\right]^{1/2},\nn\\
R_{21} &=& \pm \Delta_{(q,q')}^{-1/2} e^{\phi_0/2 + \varphi_0/{2\sqrt 3}}
\Bigg\{ e^{\sigma_0}\left[ \left(q_2 - \rho_0 q_2' \right) + \chi_{30}
\left(q_1 - \rho_0 q_1'\right) \right]^2 + e^{-\sigma_0} \left(q_2' + 
\chi_{30} q_1'\right)^2\Bigg\}^{1/2},\nn\\
R_{31} &=& \pm \Delta_{(q,q')}^{- 1/2} e^{-\varphi_0/{\sqrt 3}} \Bigg\{
e^{\sigma_0}\left[\left(q_3 - \rho_0 q_3'\right) + \chi_{10} \left(q_2 -
\rho_0 q_2'\right) + \left(\chi_{20} + \chi_{10} \chi_{30}\right)
\left(q_1 -\rho_0 q_1'\right)\right]^2\nn\\
&\,&\,\,\,\,\,\qquad\qquad + e^{-\sigma_0}\left[q_3' + 
     \chi_{10} q_2' + \left(\chi_{20} +
\chi_{10}\chi_{30}\right) q_1'\right]^2 \Bigg\}^{1/2}.
\eqaend
The scalars $\sigma$ and $\rho$ parametrizing the coset $SL(2,R)/SO(2)$
are now given by the following matrix equation
\eqabegin
{\cal M}_2 &=& A^{-\sqrt{3/28}} (\rho)\, \Lambda_{20} \left(\begin{array}{cc}
A^{\sqrt{3/7}} (\rho)&0\\
0&1\end{array}\right)\Lambda_{20}^T,\nn\\
&=&
 A^{-\sqrt{3/28}} (\rho) \,\nu_{20}\left(\begin{array}{cc}
C(\rho) \cos^2\alpha + 1&C(\rho) \cos\alpha \sin\alpha\\
C(\rho) \cos\alpha \sin\alpha&C(\rho) \sin^2\alpha + 1\end{array}\right)
\nu_{20}^T,
\eqaend
where $C(\rho) = A^{\sqrt{3/7}} (\rho) - 1$, $\nu_{20}$ is
\begineq
\nu_{20} = e^{\sigma_0/2} \left(\begin{array}{cc}
       e^{-\sigma_0}& \rho_0\\
       0&1\end{array}\right),
\endeq
and $\cos\alpha$ and $\sin\alpha$ are given by
\eqabegin
\cos\alpha &=& \pm \frac{e^{\sigma_0/2} \left(q_1 - \rho_0 q_1'\right)}{\left[
e^{\sigma_0} \left(q_1 -\rho_0 q_1'\right)^2 + e^{-\sigma_0} {q'}_1^2
\right]^{1/2}},\nn\\
\sin\alpha &=& \pm \frac{e^{-\sigma_0/2} q_1'}{\left[
e^{\sigma_0} \left(q_1 -\rho_0 q_1'\right)^2 + e^{-\sigma_0} {q'}_1^2
\right]^{1/2}}.
\eqaend
As usual, we will not present the explicit expressions for $\sigma$ and $\rho$
which can be obtained in a straightforward manner in this simple case.
The metric for the $SL(3,Z)\times SL(2,Z)$ 0-brane continues to be given by the
one in
Eq.(61) but now with $Q = Q_{(q,q')}$.
Also we expect that when any two of the three integers in each integral triplet 
are relatively prime to each other, the 0-brane is stable.

We now present the configuration for a general $SL(3, Z)\times SL(2, Z)$ 
4-brane carrying two magnetic-like integral charge triplets $p$ and $p'$.
The $\Delta$-factor is 
\begineq
\Delta_{(p, p')} = (p^T, {p'}^T) {\cal M}_{20} \left(\begin{array}{c}
{\cal M}_{30}^{- 1} p\\
{\cal M}_{30}^{- 1} p'\end{array}\right).
\endeq
The mass per unit 4-brane volume $M_{(p, p')}$, the central charge $Q_{(p, p')}$
and tension $T_{(p, p')}$ measured in Einstein frame are
\eqabegin
M_{(p, p')} &=& Q_{(p, p')} = T_{(p, p')} = \Delta_{(p, p')}^{1/2} Q_0,\nn\\
&=&\left[e^{-\sigma_0} p^T {\cal M}_{30}^{-1} p + 
e^\sigma_0 (p' + \rho_0 p)^T {\cal M}_{30}^{-1} (p' + \rho_0 p)\right]^{1/2} 
Q_0,\nn\\
&=&\Bigg(e^{-\sigma_0} \left\{e^{\phi_0 - \varphi_0/\sqrt{3}} (p_1 - \chi_{30} 
p_2 - \chi_{20} p_3)^2 + e^{-\phi_0 - \varphi_0/\sqrt{3}} 
(p_2 - \chi_{10} p_3)^2 + e^{2\varphi_0 /\sqrt{3}} p_3^2\right\}\nn\\
&\,& + e^\sigma_0\Bigg\{e^{\phi_0 - \varphi_0/\sqrt{3}}\left[ 
(p_1' + \rho_0 p_1) - \chi_{30} (p_2' + \rho_0 p_2) - \chi_{20} 
(p_3' + \rho_0 p_3)\right]^2\nn\\
&\,& + e^{-\phi_0 - \varphi_0/\sqrt{3}}\left[(p_2' + \rho_0 p_2) - \chi_{10}
(p_3' + \rho_0 p_3)\right]^2 + e^{2\varphi_0/\sqrt{3}} (p_3' + \rho_0 p_3)^2
\Bigg\}\Bigg)^{1/2} Q_0.
\eqaend

The two 2-form field strength triplets are
\begineq
\left(\begin{array}{c}
{\tilde{\cal F}}_2 \\
{\tilde F}_2\end{array}\right) =\left(\begin{array}{c}
p \\
p'\end{array}\right) Q_0 \epsilon_2.
\endeq
The metric is
\begineq
ds^2 = \left(1 + \frac{Q_{(p,p')}}{\rho}\right)^{- 1/6} \left[- (d t)^2 +
(d x^i)^2 \right] + \left(1 + \frac{Q_{(p,p')}}{\rho}\right)^{5/6}\left[
(d \rho)^2 + \rho^2 d \Omega_2^2\right],
\endeq
with $i = 1, 2, 3, 4$.

The scalars parametrizing the coset $SL(3, R)/SO(3)$ are given uniquely
 by the matrix equation
\eqabegin
{\cal M}_3 &=& A^{- 1/\sqrt{21}} (\rho)\, \Lambda_{30} \left(\begin{array}{ccc}
        A^{ \sqrt{3/7}} (\rho) & 0 &0\\
        0&1&0\\
        0&0&1\end{array}\right) \Lambda^T_{30},\nn\\
    &=& A^{- 1/\sqrt{21}} (\rho)\, 
 \nu_{30} \left(\begin{array}{ccc}
R_{11}^2 B(\rho) + 1&R_{11} R_{21} B(\rho) &R_{11} R_{31} B(\rho)\\
R_{21} R_{11} B(\rho)& 
R_{21}^2 B(\rho) + 1&
R_{21} R_{31} B(\rho)\\
R_{31} R_{11} B(\rho)& 
R_{31} R_{21} B(\rho)&
R_{31}^2 B(\rho) + 1\end{array}\right) \nu_{30}^T,
\eqaend
where $B (\rho) = A^{\sqrt{3/7}} (\rho) - 1$, and the $SO(3)$ elements are
\eqabegin
R_{11} &=& \pm \Delta_{(p, p')}^{- 1/2}\,e^{\phi_0/2 - \varphi_0/2\sqrt{3}}
\Bigg\{e^{-\sigma_0} \left(p_1 - \chi_{30} p_2 - \chi_{20} p_3\right)^2\nn\\
&\,&\,\,\,\,\, + e^\sigma_0\left[(p_1' + \rho_0 p_1) - \chi_{30} (p_2' + \rho_0 p_2)
- \chi_{20} (p_3' + \rho_0 p_3)\right]^2\Bigg\}^{1/2},\nn\\
R_{21} &=& \pm \Delta_{(p, p')}^{- 1/2}\, e^{-\phi_0/2 - \varphi_0/2\sqrt{3}}
\left\{e^{-\sigma_0} (p_2 - \chi_{10} p_3)^2 + e^\sigma_0 \left[(p_2' + \rho_0
p_2) - \chi_{10} (p_3' + \rho_0 p_3 )\right]^2\right\}^{1/2},\nn\\
R_{31} &=&\pm \Delta_{(p, p')}^{- 1/2}\,e^{\varphi_0/\sqrt{3}}\left[
e^{-\sigma_0} p_3^2 + e^\sigma_0 (p_3' + \rho_0 p_3)^2\right]^{1/2}.
\eqaend
The scalars $\sigma$ and $\rho$ parametrizing the coset $SL(2, R)/SO(2)$ are
given by the following matrix equation
\eqabegin
{\cal M}_2 &=& A^{\sqrt{3/28}} (\rho)\, \Lambda_{20} \left(\begin{array}{cc}
A^{- \sqrt{3/7}} (\rho)&0\\
0&1\end{array}\right)\Lambda_{20}^T,\nn\\
&=&
 A^{\sqrt{3/28}} (\rho) \,\nu_{20}\left(\begin{array}{cc}
C(\rho) \cos^2\alpha + 1&C(\rho) \cos\alpha \sin\alpha\\
C(\rho) \cos\alpha \sin\alpha&C(\rho) \sin^2\alpha + 1\end{array}\right)
\nu_{20}^T,
\eqaend
where $C(\rho) = A^{-\sqrt{3/7}} (\rho) - 1$, and the $\cos\alpha$ and 
$\sin\alpha$ are
\eqabegin
\cos\alpha &=& \pm \frac{e^{-\sigma_0/2} p_3}{\left[e^{-\sigma_0} p_3^2 +
e^\sigma_0 (p_3' + \rho_0 p_3)^2\right]^{1/2}},\nn\\
\sin\alpha &=& \pm \frac{e^{\sigma_0/2} (p_3' + \rho_0 p_3)}
{\left[e^{-\sigma_0} p_3^2 + e^\sigma_0 (p_3' + \rho_0 p_3)^2\right]^{1/2}}.
\eqaend
In the above
\begineq
A (\rho) = \left(1 + \frac{Q_{(p,p')}}{\rho}\right)^{\sqrt{7/3}}.
\endeq

We also expect as before that when any two of the three integers 
in either of the 
two integral charge triplets are relatively prime, then the 4-brane is
stable. This completes the constructions of all the p-brane solutions
in $D = 8$ type II string theory.

\section{U-duality p-Branes}

The previous sections along with the previous studies [9,11,12] 
lay out the 
ground for us to construct the general U-duality p-branes of various 
supergravity theories in diverse dimensions.
To avoid possible complicaions, we limit ourselves to $10 \ge D \ge 4$. We also
set the restriction that both $p \ge 0$ and $D - p - 4 \ge 0$, i.e., the 
spatial dimensions of both a p-brane and its magnetic dual $(D - p - 4)$-brane 
in $D$ dimensions are greater than or equal to zero. The reason for this latter 
limitation is that in order to realize the classical 
Cremmer-Julia symmetries in
$D \le 7$ either at the level of action or at the level of equation of 
motion (EOM),  
every field 
strength should be dualized whenever this results in a field strength of a 
smaller degree, as pointed out in [16,15]. 
  
It apperas that we have two cases to study, depending on whether the 
Cremmer-Julia symmetry is realized naturally at the level of supergravity 
action or EOM. The latter 
consists of the possible dyonic objects, i.e., membranes in $D = 8$, strings 
in $D = 6$ and 0-branes in $D = 4$.  Note that the dyonic solutions have some
crucial differences from their non-dyonic counterparts. For example, the 
classical 
Cremmer-Julia symmetries associated with the dyonic objects break into the 
corresponding U-duality symmetries due to instanton 
effects rather than  the charge quantizations as for the $SL(3, Z)$ strings
and for all other U-duality non-dyonic p-branes.
Furthermore, the `electric'-charge carried by a dyonic object 
is in general not 
quantized integrally. A special example of the construction of U-duality dyonic 
membranes is given in [12]. But once a $p$-`magnetic' charge and 
a $q$-`electric' charge, which satisfy the corresponding dyonic 
quantization rule,
are assigned to a dyonic object, we can employ the property of the maximal 
compact group of the corresponding Cremmer-Julia symmetry to determine the 
corresponding $\Delta$-factor in terms of $p$ and $q$ and the vacuum moduli 
as we did for the $SL(3, R)$ strings in section 4. This in turn 
determines the corresponding central charge, ADM mass per unit p-brane volume 
and tension. Therefore, the construction of U-duality dyonic objects 
from a given
solution is not much different from that of the non-dyonic p-branes. 
Actually, even
for each of the dyonic cases, we can also realize the corresponding 
Cremmer-Julia symmetry formally at the level of action by introducing a second 
set of field strengths but with a constraint imposed at the level of EOM as 
discussed recently in [15]. These field strengths together 
with the original 
ones form a certain representation of the Cremmer-Julia symmetry. Such a formal
action is useful for our unifying discussion of U-duality p-branes.

The study of the $SL(3, Z)$ strings (3-branes) and the $SL(3,Z)\times SL(2, Z)$
 0-branes (4-branes)
here alongwith the previous examples, i.e.,
the $SL(2,Z)$ strings [9] and fivebranes [11] as well as the $D = 8$ 
dyonic membranes [12], indicates that we really do not need to go 
through the whole procedure to 
construct these solutions as we did for the $SL(3,Z)$ strings 
in section 4.  We
can simply write down the these solutions  in each case as we did for 
the $SL(3, Z)\times  SL(2, Z)$ 0-branes (4-branes) in the previous section, 
if we know the scalar coset matrix ${\cal M}$, 
the transformation law of all the relevant field strengths (or gauge potentials)
under the corresponding Cremmer-Julia symmetry, and a simple p-brane  
solution carrying a single charge associated with one 
of the field strengths (we
always choose that field strength as the first component of the corresponding 
column vector). 
The other important quantity for the construction of solutions is the 
$\Delta$-factor which can be deduced quite easily  
as it has to remain invariant under the corresponding U-duality symmetry. 
By employing the procedure just outlined we will construct here the U-duality
p-brane solutions of various supergravity theories in diverse dimensions. 
Before we do so, we like to discuss certain general properties of 
various supergravity theories which will facilitate our constructions 
for these solutions.

For every supergravity theory in $D \le 10$, there exist a non-compact global
symmetry $G$ realized non-linearly and a hidden compact local symmetry $H$ of 
the theory [2,16]. The scalars in the theory are always 
described by the coset $G/H$.
$H$, which is isomorphic to the maximal compact subgroup of $G$, is the 
automorphism symmetry of the algebra of supersymmetry. The tensor fields in 
the theory always form certain representations of $G$ but inert under $H$. On 
the other hand, the spinors always transform under $H$ wihle they are inert 
under $G$. Therefore, transformations of $G$ preserve supersymmetry. This 
immediately implies that any new p-brane solution obtained by a transformation 
of $G$ on a given p-brane solution will preserve the same number of unbroken
 supersymmetries as the original p-brane does. The central charge associated 
with these solutions should also be invariant under $G$.
In Einstein frame, the metric is always a singlet of $G$. Therefore, 
the ADM mass per unit p-brane volume is also a singlet of $G$. We must conclude
that the $\Delta$-factor for any p-brane should also be invariant in general 
under $G$ which is consistent with our observation for all the specific examples
studied so far. To find a general static p-brane solution, the spinors as well
as all the gauge potentials except the $(p +1)$-form ones in a 
supergravity are always set to zero. Under such a circumstance, we can always
choose the scalar coset matrix ${\cal M}$ to transform in the same 
representation of the non-compact group $G$ as the $(p+1)$-form gauge 
potentials or $(p + 2)$-form field strengths do.  This feature is useful
for our subsequent discussions.  It should be noted that given $G$ as a 
symmetry of a supergravity
theory, it is sufficient to consider the lowest order bosonic action
involving the graviton, the scalars parametrizing the coset $G/H$, and the
$(p+2)$-form field strengths
to determine how the scalar coset matrix ${\cal M}$ and the column vector 
$A_{p + 1}$ of the $(p + 1)$-form gauge potentials $A_{p + 1}^{(i)}$  
transform.  
     
The lowest-order bosonic action associated with non-dyonic U-duality 
electric-like p-branes or magnetic-like $(D - p - 4)$-branes  
in $D$ dimensions can be cast in the Einstein frame as
\begineq
S_D = \int\, d^Dx \sqrt{- g} \left[R + \frac{1}{4}\, {\rm tr} 
\nabla_\mu {\cal M} \nabla^\mu {\cal M}^{ -1} - \frac{1} {2 (p + 2) !} 
{\tilde F}^T_{(p + 2)} {\cal M} {\tilde F}_{(p + 2)}\right],
\endeq
where  the column vector ${\tilde F}_{(p + 2)}$ is defined as
\begineq
{\tilde F}_{(p + 2)} = \left(\begin{array}{c}
                              d A_{(p + 1)}^{(1)}\\
                              d A_{(p + 1)}^{(2)}\\
                              \cdot\\   
  				\cdot\\
  				\cdot\end{array}\right).
\endeq
The above action is invariant under the following transformation
\begineq
g_{\mu\nu} \rightarrow g_{\mu\nu},\,\, {\cal M}\rightarrow \Lambda {\cal M} 
\Lambda^T,\,\, F_{(p + 2)} \rightarrow (\Lambda^T)^{-1} F_{(p + 2)},
\endeq
where $\Lambda$ is a global Cremmer-Julia symmetry matrix. 

Here we present only U-duality p-brane solutions preserving half of the 
spacetime supersymmetry. U-duality p-brane solutions preserving less than half 
of the spacetime supersymmetry as well as U-duality black p-branes can be 
written down in exactly the same fashion. As discussed earlier, 
the number of unbroken
supersymmetries associated with a U-duality p-brane is completely determined 
by that of the initial NSNS p-brane configuration. 

The NSNS $(d - 1)$-brane or $({\tilde d} - 1)$-brane configuartions in 
diverse dimensions have been obtained some time ago in [6]. For a 
$(d - 1)$-brane carrying electric-like
charge $Q (d)$, we have, for zero asymptotic value of the dilaton,
\eqabegin
&&ds^2 = \left(1 + \frac{Q (d)}{{\tilde d}\, \rho^{\tilde d}}
\right)^{- {\tilde d}/(d + {\tilde d})} \left[- (d t)^2 + (d x^i)^2\right] +  
\left(1 + \frac{Q (d)}{{\tilde d}\, \rho^{\tilde d}}\right)^{d/(d + 
{\tilde d})} \left[(d\rho)^2 + \rho^2 d \Omega_{{\tilde d} + 1}^2\right],\nn\\
&&e^{ - 2 \Phi} = \left(1 + \frac{Q (d)}{{\tilde d}\, \rho^{\tilde d}}
\right)^{\alpha(d)} = A_d (\rho),\,\,\, {\tilde F}_{d + 1}^{(1)} = Q (d)\, 
A^{- \alpha(d)/2}_d (\rho) \,\ast \epsilon_{{\tilde d} + 1},
\eqaend
where $d = p + 1, {\tilde d} = D - p -3$, $d \Omega_n^2$ is the
metric on the unit $n$-sphere and $\epsilon_n$ is the corresponding volume
form.  The magnetic dual of the above $(d - 1)$-brane, i.e., the 
$({\tilde d} - 1)$-brane, can be obtained from the above by the following 
replacements: $e^{- \alpha (d) \Phi} \ast {\tilde F}_{d + 1}^{(1)} \rightarrow 
{\tilde F}_{{\tilde d} + 1}^{(1)}$, $\Phi \rightarrow - \Phi$, 
$d \leftrightarrow {\tilde d}$. They are solutions of the following action
\begineq
S_D = \int d^D\,x \sqrt{- g}\left[ R - \frac{1}{2} (\partial \Phi)^2 -
\frac{1}{2 (d + 1)!}\, e^{ - \alpha (d) \Phi} \, 
({\tilde F}_{d + 1}^{(1)})^2\right].
\endeq
In the above, the parameter $\alpha (d)$ is given by
\begineq
\alpha (d) = \sqrt{4 - \frac{2 d {\tilde d}}{d + {\tilde d}}}.
\endeq
 
If we denote the electric-like 
integral charge column vector associated with a general U-duality 
$(d - 1)$-brane as 
$q$ and the magnetic-like integral charge column vector associated with a 
general U-duality $({\tilde d} -1)$-brane as $p$, the U-duality invariant 
$\Delta$-factor is, for the $(d - 1)$-brane,
\begineq 
\Delta_q (d) = q^T {\cal M}_0^{- 1} q,
\endeq
and, for the $({\tilde d} - 1)$-brane
\begineq
\Delta_p ({\tilde d}) = p^T {\cal M}_0 \,p.
\endeq

The ADM mass per unit $(d - 1)$-brane volume $M_q (d)$, the 
central charge $Q_q (d)$ and the
$(d - 1)$-brane tension $T_q (d)$ measured in Einstein metric are
\begineq
M_q (d) = Q_q (d) = T_q (d) = \Delta^{1/2}_q Q_0 (d),
\endeq
where $Q_0 (d)$ is the unit charge which can be taken as the fundamental 
$(d - 1)$-brane tension. 
The same relations for $({\tilde d} - 1)$-brane can be obtained from 
the above by the following replacements: $q \rightarrow p$, $d \rightarrow 
{\tilde d}$. The field strength column vector ${\tilde F}_{d + 1}$ for the 
$(d - 1)$-brane is now
\begineq
{\tilde F}_{d + 1} = {\cal M}_0^{-1} \,q\, Q_0 (d)\, A_d^{- \alpha (d)/2} (\rho)\,
 \ast \epsilon_{{\tilde d} + 1},
\endeq
while the field strength column vector ${\tilde F}_{d + 1}$ for the 
$({\tilde d} - 1)$-brane is
\begineq
{\tilde F}_{d + 1} = p\, Q_0 ({\tilde d}) \epsilon_{d + 1}.
\endeq 

The metric for the $(d - 1)$-brane is still given by the one in Eq.(87) but now 
with $Q (d) = Q_q (d)$.
The same is true for the $({\tilde d} -1)$-brane but now with $Q ({\tilde d}) =
Q_p ({\tilde d})$. 

The scalars for either the $(d - 1)$-brane or 
the $({\tilde d} - 1)$-brane can be 
determined uniquely by the matrix equation
\begineq
{\cal M} = \Lambda_0 \,{\cal M}_{\rm initial}\, \Lambda_0^T,
\endeq
where ${\cal M}_{\rm initial}$ is the scalar coset matrix describing
the initial NSNS $(d - 1)$-brane or 
$({\tilde d} - 1)$-brane configuration. We expect 
that ${\cal M}_{\rm initial}$ approaches unity as $\rho \rightarrow \infty$.
It is described by the function $A_d (\rho)$ for the $(d - 1)$-brane or 
the $A_{\tilde d} (\rho)$ for $({\tilde d} - 1)$-brane and  has 
only non-vanishing diagonal elements. As for the case of $SL(3, Z)$ strings, 
the  above
Cremmer-Julia symmetry matrix $\Lambda_0$ is only partially determined by
the equation ${\cal M}_0 = \Lambda_0 \Lambda_0^T$. Nevertheless, the scalar
coset matrix ${\cal M}$ is completely determined  in terms of 
the asymptoptic values of the 
scalars, the charge $q$ (or $p$) and the function $A_d (\rho)$ (or 
the function $A_{\tilde d} (\rho)$) as we have seen in our previous examples.
Also note that ${\cal M}$ should approach 
${\cal M}_0$ as 
$\rho \rightarrow \infty$.

We now come to discuss the dyonic U-duality p-branes.
 For each of the dyonic p-branes with $p = 0, 1, 2$ ($D = 2p + 4 = 4, 6, 8$), 
the $(p + 1)$-form gauge potentials appear in the corresponding action only
through their $(p + 2)$-form field strengths. These field strengths 
${\tilde F}_{p + 2}^{(i)}$ by 
themselves  do not form a representation of the corresponding Cremmer-Julia
symmetry.  However, if we introduce  an equal number of $p + 2$-form field 
strengths ${\tilde G}_{p + 2}^{(i)}$, then the column vector 
${\tilde H}_{p + 2}$ 
\begineq
{\tilde H}_{p + 2} = \left(\begin{array}{c}
                  {\tilde F}_{p + 2}\\
                  {\tilde G}_{p + 2}\end{array}\right),
\endeq
does form a fundamental representation of the corresponding Cremmer-Julia 
symmetry group $G$. The lowest-order bosonic action in Einstein frame can now
 be expressed formally as
\begineq
S_{2 p + 4} = \int\,d^{2p + 4} x \sqrt{- g} \left[R + \frac{1}{4}\,{\rm tr}
\,\nabla_\mu {\cal M} \nabla^\mu {\cal M}^{-1} - 
\frac{1}{4 (p + 2)!}\, {\tilde H}^T_{p + 2}{\cal M} {\tilde H}_{p + 2}\right],
\endeq
which is invariant under the following transformations
\begineq
g_{\mu\nu} \rightarrow g_{\mu\nu},\,\, {\cal M}\rightarrow \Lambda {\cal M}
\Lambda^T,\,\, {\tilde H}_{p + 2} \rightarrow (\Lambda^T)^{-1} 
{\tilde H}_{p + 2}
\endeq
with $\Lambda$ the Cremmer-Julia symmetry group matrix. The equations of motion
from the above action reduce to the original ones only as the 
covariant constraint relation ${\tilde H}_{p + 2} = 
\Omega\, {\cal M} \ast {\tilde H}_{p + 2}$ is 
imposed at the level of EOM as
discussed recently in [15] with $\Omega$ the invariant matrix of the 
corresponding Cremmer-Julia symmetry group. It is given by
\begineq
\Omega = \left(\begin{array}{cc}
               0& (- 1)^{p + 1} I\\
  		I&0\end{array}\right),
\endeq
where $I$ is the unit matrix. With the constraint imposed, the 
Bianchi identity $d {\tilde G}_{p + 2} = 0$ is actually the original 
equation of motion
for the field strength ${\tilde F}_{p + 2}$. So $d {\tilde H}_{p + 2} = 0$ 
gives rise to a charge vector ${\cal Z}$, i.e.,
\begineq
{\cal Z} = \left(\begin{array}{c} 
           p\\ q\end{array}\right),
\endeq
where $p$ and $q$ correspond respectively to the magnetic and electric 
charge column 
vectors associated with the dyonic p-brane, i.e.,
\begineq
p = \frac{1}{V_{p + 2}} \int_{S^{p + 2}} {\tilde F}_{p + 2},\,\,\, 
q = \frac{1}{V_{p + 2}} \int_{S^{p + 2}} {\tilde G}_{p + 2},
\endeq
with $S^{p + 2}$ the asymptotic $(p + 2)$-sphere, and $V_{p + 2}$ the volume of 
unit $(p + 2)$-sphere. Two of such charge vectors
${\cal Z}$ and ${\cal Z}'$ obey the dyonic quantization rule
\begineq
{\cal Z}^T \Omega {\cal Z}' = q^T p' + ( - 1)^{p + 1} p^T q' \in Z.
\endeq

We like to emphasize that the above formal action only serves us the purpose  
to identify the scalar coset matrix, to deduce the transformations as 
given above, and to draw analogy with the non-dyonic cases discussed above.
For the construction of the dyonic U-duality p-branes, we only employ those
 transformation relations but not the action. However, the constraint relation 
is always imposed. In other words, $d {\tilde G}_{p + 2} = 0$ is 
the equation of motion for the field strength ${\tilde F}_{p + 2}$.      

Starting with a NSNS p-brane configuration carrying a pure magnetic charge 
$Q (p + 1)$ in $D = 2 p + 4$ as described right after Eq.(87), we can obtain a 
general U-duality dyonic p-brane in a similar way as we did for the U-duality
non-dyonic p-branes carrying magnetic charges. 
 The corresponding $\Delta$-factor is now
\begineq
\Delta_{\cal Z} (p + 1) = {\cal Z}^T {\cal M}_0 {\cal Z},
\endeq
which is invariant under the corresponding U-duality transformation. Then
the ADM mass per unit dyonic p-brane volume $M_{\cal Z} (p + 1)$ and the 
central charge $Q_{\cal Z} (p + 1)$ are
\begineq
M_{\cal Z} (p + 1) = Q_{\cal Z} (p + 1) = \Delta_{\cal Z}^{1/2}\, Q_0 (p + 1),
\endeq
where $Q_0 (p + 1)$ is again the unit of charge which can be taken as 
the fundamental
NSNS p-brane tension. The metric remains the same as that for 
the initial NSNS p-brane but now with $Q (p + 1) = Q_{\cal Z} (p + 1)$. 
For $D = 2 p + 4$, the form of metric is much simpler since $d = {\tilde d}$.
The scalars can be obtained exactly the same way as for non-dyonic p-branes
carrying magnetic charges. But there is an important difference in determining
the field strengths for the U-duality dyonic p-brane. We start with
${\tilde F}_{p + 2}^{(1)} = Q_{\cal Z} (p + 1)\, \epsilon_{p + 2}$. 
If we impose the constraint ${\tilde H}_{p + 2} = \Omega\, {\cal M}\,
\ast {\tilde  H}_{p + 2}$ to 
obtain ${\tilde G}^{(1)}_{p + 2}$ from the outset, then the constraint will be 
automatically satisfied for the U-duality dyonic p-brane. 
Therefore, taking $d {\tilde G}^{(1)}_{p + 2} = 0$ as just the 
equation of motion for ${\tilde F}^{(1)}_{p + 2}$, we have 
${\tilde G}^{(1)}_{p + 2} = Q_{\cal Z} (p + 1)\, A_{p + 1}^{- \sqrt{3 - p} /2} 
(\rho)\, \ast \epsilon_{p + 2}$ with 
\begineq
A_{p + 1} (\rho) = \left(1 + \frac{Q_{\cal Z} (p + 1)}{(p + 1)\, \rho^{p + 1}}
\right)^{\sqrt {3 - p}}.
\endeq
With the above ${\tilde F}^{(1)}_{p + 2}$ and ${\tilde G}^{(1)}_{p + 2}$, we 
have the
${\tilde H}_{p + 2}$ for the U-duality dyonic p-brane as
\eqabegin
{\tilde H}_{p + 2} &=& \left(\begin{array}{c}
                    {\tilde F}_{p + 2}\\
                    {\tilde G}_{p + 2}\end{array}\right)\nn\\
         &=& \left[ \left(\begin{array}{c}
                    p\\
                    q\end{array}\right) \epsilon_{p + 2} + \Omega {\cal M}_0
\left(\begin{array}{c}
                    p\\
                    q\end{array}\right) A_{p + 1}^{- \sqrt{3 - p} /2} (\rho)\,
\ast\epsilon_{p + 2}\right] Q_0 (p + 1).
\eqaend  

We have now completed our constructions of both the dyonic and the
non-dyonic U-duality p-brane solutions 
in diverse dimensions. In order to see how
various quantities depend on the asymptotic values of the scalars, we have to
give an explicit parametrization of the coset matrix ${\cal M}$ in terms of 
these scalars. This can be done without much difficulty based on various 
known supergravity theories in diverse dimensions. Further, to see how these 
quantities depend on the string
coupling constant and the asymptotic values for various scalars and axions,
 we have to follow the route, described in section 3, to construct the coset 
matrix ${\cal M}$. In general, this must be very tedious. But in principle,
it can always be done. 
Without the explicit form for ${\cal M}$, we can still, for 
example, give the criteria for the stability of the U-duality p-branes. For a
non-dyonic p-brane carrying either an electric-like or a magnetic-like 
integral charge column vector, this p-brane is absolutely stable if any two 
integers in the corresponding charge column vector are relatively prime.
For a dyonic p-brane, the magnetic-like charge column vector can in general 
be integral but the electric-like charge column vector cannot as discused in 
[12]. Nevertheless the dyonic p-brane is still stable if any two integers in
the magnetic-like integral charge column vector are relatively prime.
As for the case of $SL(3,Z)$-string, the general U-duality p-brane solution 
contains
all the information about the corresponding U-duality p-brane multiplets. 
Other similar discussions can also be made here as we did for the $SL(3,Z)$ 
strings. In what follows, we will give a brief discussion about 
possible U-duality p-branes in each of the $10 \ge D \ge 4$ maximal 
supergravity theories.              

\begin{itemize}

\item In $D = 10$ type IIB supergravity, there is a well-known 
Cremmer-Julia type 
symmetry $SL(2,R)$. The corresponding U-duality symmetry is conjectured 
to be $SL(2,Z)$. In this theory, there are  two 2-form potentials forming a 
doublet of $SL(2,R)$, one 4-form potential which is a singlet of $SL(2,R)$ and
whose field strength is self-dual, and two scalars parametrizing the coset
$SL(2,R)/SO(2)$. We therefore expect $SL(2,Z)$ superstrings [9] and $SL(2,Z)$
superfivebranes [11] both of which were constructed recently. The self-dual 
3-brane was found sometime ago [8,25].

\item Maximal supergravity in $D = 9$ has a Cremmer-Julia symmetry 
$GL(2,R) \simeq SL(2,R) \times SO(1,1)$. The corresponding conjectured U-duality
symmetry is $SL(2,Z) \times Z_2$ with $SL(2,Z)$ a strong-weak duality symmetry
and $Z_2$ a T-duality symmetry. In this theory, there are a dilaton inherited 
from $D = 10$ and an axion parametrizing the coset $SL(2,R)/SO(2)$, a second 
dilatonic scalar which is a singlet of the $SL(2,R)$, one 3-form gauge 
potential which is also a singlet of the $SL(2,R)$, two 2-form gauge potentials
forming a doublet of the $SL(2,R)$, and three 1-form gauge potentials 
two of which
forms a doublet of the $SL(2,R)$ while the other is a singlet. The $SO(1,1)$ 
symmetry transforms the second dilatonic scalar by a constant shift and rescales
all the other gauge potentials but leaves the $D = 10$ dilaton and the axion 
inert. That is why the $SO(1,1)$ is merely a classical T-duality symmetry. This
symmetry will not be useful in generating new solutions. The bosonic action 
exhibiting the $SL(2,R) \times SO(1,1)$ symmetry has been given explicitly, for
example, in [26]. The scalar matrix ${\cal M}$ parametrizing the coset 
$SL(2,R)/SO(2)$ is a familiar one. The $SL(2,Z)$ strings carrying electric-like 
charges and $SL(2,Z)$ 4-branes carrying magnetic-like charges can be read 
off readily from our general formulae. There are also $SL(2,Z)$ 0-branes and 
$SL(2,Z)$ 5-branes.  There are also solutions which are inert under 
the $SL(2,Z)$,
namely, 2-brane and 3-brane as well as 0-brane and 5-brane found some time ago 
in [6].

\item The various U-duality p-branes 
in $D = 8$ maximal supergravity have all been 
studied in this paper and in [12]. As discussed in detail 
in the previous sections, we have $SL(2,Z)$
dyonic membranes, $SL(3,Z)$ strings and 3-branes and $SL(3,Z) \times SL(2,Z)$
0-branes and 4-branes.

\item The Cremmer-Julia symmetry 
in $D = 7$ maximal supergravity is $SL(5,R)$. The 14
scalars in this theory parametrize the coset $SL(5,R)/SO(5)$. The
conjectured U-duality symmetry is $SL(5,Z)$.  There are five 2-form gauge 
potentials forming 5-dimensional fundamental representation of $SL(5,R)$ and
ten 1-form gauge potentials forming a 10-dimensional irreducible representation
of $SL(5,R)$. We have therefore $SL(5,Z)$ strings and membranes for which the
$5 \times 5$ scalar coset matrix ${\cal M}$ can be found in [2,15]. 
We also have
$SL(5,Z)$ 0-branes and 3-branes for which the $10 \times 10$ scalar coset 
matrix ${\cal M}$ can be parametrized explicitly based on the known
$D = 7$ supergravity [27].

\item Maximal supergravity in $D = 6$ has a 
Cremmer-Julia symmetry $SO(5,5)$. The
25 scalars appearing in this theory parametrize the coset 
$SO(5,5)/SO(5) \times SO(5)$. The conjectured U-duality symmetry is $SO(5,5;Z)$.
There are five 2-form gauge potentials which appear in the action only through
their 3-form field strengths. This is the dimension for which dyonic string 
solutions appear. These five 3-form field strengths do not form a 
representation of $SO(5,5)$ but as discussed at length before, this symmetry
can be realized at the level of EOM through interchanging the Bianchi 
identities and the equations of motion for the 3-form field strengths. In other
words, the five equations and five Bianchi identities do form a 10-dimentional
fundamental representation of $SO(5,5)$. We therfore have $SO(5,5;Z)$ dyonic
strings for which the $10 \times 10$ scalar coset matrix ${\cal M}$ has been 
constructed, for example, in [28,15]. There are also sixteen 1-form gauge 
potentials which in this case form a 16-dimensional spinor representation of
$SO(5,5)$. We have also $SO(5,5;Z)$ 0-branes and membranes for which the
$16\times 16$ scalar coset matrix ${\cal M}$ can be parametrized explicitly
based on the already known $D = 6$ supergravity theory [29].

\item The Cremmer-Julia symmetry in $D = 5$ is the non-compact $E_{6(+6)}$. 
There are
42 scalars in this theory which parametrize the coset $E_6/USp(8)$. The 
conjectured U-duality symmetry is $E_6 (Z)$. In this theory, there are only 
twenty seven 1-form gauge potentials which form a 27-dimensional fundamental 
representation of $E_6$. Therefore, we have $E_6 (Z)$ 0-branes and strings for
which the $27 \times 27$ scalar coset matrix ${\cal M}$ can be parametrized
explicitly based on the well-studied $D = 5$ supergravity theory [30].

\item Maximal supergravity in $D = 4$ has a Cremmer-Julia symmetry $E_{7(+7)}$.
The seventy scalars in this theory parametrize the coset $E_{7(+7)}/SU(8)$. The
conjectured U-duality symmetry is $E_7 (Z)$.  This is the dimension for which
we have dyonic 0-branes. There are only twenty eight 
1-form gauge potentials which appear in the action only through their 2-form
field strengths. Similar to the cases in $D = 6$ and $D = 8$ for dyonic
strings and dyonic membranes, these twenty eight 2-form field strengths 
combined  with the other twenty eight 2-form field strengths whose Bianchi 
identities give the 28 equations of motion form a 56-dimensional fundamental 
representation of $E_{7(+7)}$. We have therefore $E_7 (Z)$ dyonic 0-branes for
which the $56 \times 56$ scalar coset matrix ${\cal M}$ is given in [1,16].

\end{itemize}

\vfil\eject
\begin{Large}
\begin{center}
{\bf Appendix}
\end{center}
\end{Large}

In this appendix, we will present the field strengths without tildes
 in $D = 8$ maximal supergravity (i.e., those discussed in section 3), obtained 
directly by a $T^2$ compactification of $D = 10$ type IIA 
supergravity theory. In our notations, we have
\eqabegin
F_4 &=& {\tilde F}_4 - {\tilde F}_3^{(1)} \wedge {\cal A}_1^{(1)}  
       - {\tilde F}_3^{(2)} \wedge{\cal A}_1^{(2)}
       - {\tilde F}_3^{(3)} \wedge {\cal A}_1^{(3)}\nn\\
&\,& - \chi_1 {\tilde F}_3^{(2)} \wedge {\cal A}_1^{(3)}
      - \left(\chi_2 + \chi_1 \chi_3\right) 
        {\tilde F}_3^{(1)} \wedge  {\cal A}_1^{(3)}
     - \chi_3 {\tilde F}_3^{(1)} \wedge {\cal A}_1^{(2)} \nn\\
&\,& + {\tilde F}_2^{(3)} \wedge {\cal A}_1^{(1)}\wedge
     {\cal A}_1^{(2)}
     - {\tilde F}_2^{(2)} \wedge{\cal A}_1^{(1)}\wedge
     {\cal A}_1^{(3)}
     + {\tilde F}_2^{(1)} \wedge  {\cal A}_1^{(2)}\wedge
     {\cal A}_1^{(3)}\nn\\
&\,& + \chi_1 {\tilde F}_2^{(3)} \wedge {\cal A}_1^{(1)}\wedge
     {\cal A}_1^{(3)}
     + \chi_2 {\tilde F}_2^{(3)} \wedge {\cal A}_1^{(3)}\wedge
     {\cal A}_1^{(2)}
     - \chi_3 {\tilde F}_2^{(2)} \wedge {\cal A}_1^{(2)}\wedge
     {\cal A}_1^{(3)}\nn\\
&\,& - d\rho \wedge {\cal A}_1^{(1)} 
     \wedge {\cal A}_1^{(2)}\wedge {\cal A}_1^{(3)}.
\eqaend

\eqabegin
F_3^{(1)} &=&{\tilde F}_3^{(1)} 
            + {\tilde F}_2^{(3)}\wedge {\cal A}_1^{(2)}
            -{\tilde F}_2^{(2)}\wedge{\cal A}_1^{(3)}
            + \chi_1 {\tilde F}_2^{(3)}\wedge{\cal A}_1^{(3)}
            + d \rho \wedge {\cal A}_1^{(2)} \wedge {\cal A}_1^{(2)},\nn\\
F_3^{(2)} &=& {\tilde F}_3^{(2)} + \chi_3 {\tilde F}_3^{(2)} 
              -  {\tilde F}_2^{(3)}\wedge{\cal A}_1^{(1)}
               + {\tilde F}_2^{(1)}\wedge{\cal A}_1^{(3)}
           - \chi_3 {\tilde F}_2^{(2)}\wedge{\cal A}_1^{(3)}
               -{\tilde F}_2^{(3)}\wedge{\cal A}_1^{(3)}\nn\\
          &\,&  - d \rho \wedge {\cal A}_1^{(1)}\wedge {\cal A}_1^{(3)},\nn\\
F_3^{(3)} &=& {\tilde F}_3^{(3)}  + \chi_1 {\tilde F}_3^{(2)}        
             + \left(\chi_2 + \chi_1 \chi_3\right) {\tilde F}_3^{(1)}
           + {\tilde F}_2^{(2)}\wedge{\cal A}_1^{(1)} - 
               {\tilde F}_2^{(1)}\wedge{\cal A}_1^{(2)}\nn\\ 
          &\,&    -\chi_1 {\tilde F}_2^{(3)}\wedge{\cal A}_1^{(1)}
              + \chi_2 {\tilde F}_2^{(3)}\wedge{\cal A}_1^{(2)}
          + \chi_3 {\tilde F}_2^{(2)}\wedge{\cal A}_1^{(2)}\nn\\
          &\,&    + d \rho \wedge {\cal A}_1^{(1)}\wedge {\cal A}_1^{(2)},
\eqaend

\eqabegin
F_2^{(1)} &=& {\tilde F}_2^{(1)}  - \chi_3 {\tilde F}_2^{(2)} 
             - \chi_2 {\tilde F}_2^{(3)} - d \rho \wedge {\cal A}_1^{(1)},\nn\\
F_2^{(2)} &=& {\tilde F}_2^{(2)} - \chi_1 {\tilde F}_2^{(3)} 
             - d \rho \wedge {\cal A}_1^{(2)},\nn\\
F_2^{(3)} &=& {\tilde F}_2^{(3)} - d \rho \wedge {\cal A}_1^{(3)}.
\eqaend

\eqabegin
{\cal F}_2^{(1)} &=& {\tilde {\cal F}}_2^{(1)} + 
                 d \chi_3 \wedge {\cal A}_1^{(2)} + 
     \left(d \chi_2  + \chi_1 d \chi_3\right) \wedge {\cal A}_1^{(3)},\nn\\
{\cal F}_2^{(2)} &=& {\tilde F}_2^{(2)} + d \chi_1 \wedge {\cal A}_1^{(3)},
\nn\\
{\cal F}_2^{(3)}&=& {\tilde {\cal F}}_2^{(3)}.
\eqaend
\vfil\eject
\begin{Large}
\noindent{\bf Acknowledgements}
\end{Large}

\medskip

We are grateful to M. J. Duff and E. Sezgin for discussions and to
J. H. Schwarz for an e-mail correspondence. JXL acknowledges the support
of NSF Grant PHY-9722090.

\bigskip

\begin{Large}
\noindent{\bf References}
\end{Large}

\medskip

\begin{enumerate}

\item E. Cremmer and B. Julia, \pl 80 (1987) 48; \np 156 (1979) 141;
 B. Julia, in {\it Superspace and Supergravity}, Eds. S. W.
Hawking and M. Rocek (Cambridge University Press), 1981.
\item M. J. Duff and J. X. Lu, \np 347 (1990) 394.
\item C. M. Hull and P. K. Townsend, \np 294 (1995) 196.
\item A. Giveon, M. Porrati and E. Rabinovici, Phys. Rep. C244 (1994)
77; E. Alvarez, L. Alvarez-Gaume and Y. Lozano, {\it An Introduction
to T-Duality in String Theory}, hep-th/9410237.
\item A. Sen, Int. Jour. Mod. Phys. A9 (1994) 3707.
\item M. J. Duff, R. R. Khuri and J. X. Lu, Phys. Rep. C259 (1995) 213; 
M. J. Duff and J. X. Lu, \np 416 (1994) 301.
\item A. Dabholkar, G, Gibbons, J. A. Harvey and F. Ruiz-Ruiz, \np 340
(1990) 33; A. Dabholkar and J. A. Harvey, Phys. Rev. Lett. 63 (1989) 478.
\item G. T. Horowitz and A. Strominger, \np 360 (1991) 197.
\item J. H. Schwarz, \pl 360 (1995) 13 (hep-th/9508143).
\item S. Roy, \pl 421 (1998) 176.
\item J. X. Lu and S. Roy, {\it An $SL(2, Z)$ Multiplet of Type IIB Super
Five-Branes}, hep-th/9802080 (to appear in Phys. Lett. B).
\item J. M. Izquierdo, N. D. Lambert, G. Papadopoulos and P. K. Townsend,
\np 460 (1996) 560.
\item H. Lu and C. Pope, \np 465 (1996) 127; J. P. Gauntlett, 
G. W. Gibbons, G. Papadopoulos and P. K. Townsend, \np 500 (1997) 133. 
\item R. I. Nepomechie, Phys. Rev. D31 (1985) 1921;
C. Teitelboim, \pl 167 (1986) 69.
\item E. Cremmer, B. Julia, H. Lu and C. Pope, {\it Dualisation of Dualities
I}, hep-th/9710119.
\item J. Scherk and J. Schwarz, \np 153 (1979) 61; E. Cremmer, in
{\it Supergravity 1981}, Eds. S. Ferrara and J. G. Taylor (Cambridge
University Press, 1982).
\item E. Kiritsis and B. Pioline, \np 508 (1997) 509; N. Berkovits and C.
Vafa, {\it Type IIB $R^4\,H^{4G - 4}$ Conjectures}, hep-th/9803145.
\item J. X. Lu and S. Roy, in preparation.
\item A. Font, L. Ibanez, D. Lust and F. Quevedo, \pl 249 (1990) 35;
S. J. Rey, Phys. Rev. D43 (1991) 526; A. Shapere, S. Trivedi, and F. Wilczek,
Mod. Phys. Lett. A6 (1991) 2677; M. J. Duff and R. R. Khuri,
\np 411 (1994) 473.
\item J. Polchinski, Phys. Rev. Lett. 74 (1995) 4724.
\item M. J. Duff, P. Howe, T. Inami and K. S. Stelle, \pl 191 (1987)
70.
\item J. H. Schwarz, {\it Lectures on Superstring and M-theory
Dualities}, hep-th/9607201.
\item M. J. Duff and J. X. Lu, \np 354 (1991) 141.
\item C. G. Callan, J. A. Harvey and A. Strominger, \np 359 (19910 611.
\item M. J. Duff and J. X. Lu, \pl 273 (1991) 409.
\item E. Bergshoeff, C. M. Hull and T. Ortin, \np 451 (1995) 547; H. Lu,
C. N. Pope and K. S. Stelle, \np 476 (1996) 89; J. Maharana, \pl 402
(1997) 64; S. Roy, {\it On S-Duality of Toroidally Compactified Type IIB
String Effective Action}, hep-th/9705016 (to appear in Int. Jour. Mod.
Phys. A). 
\item E. Sezgin and A. Salam, \pl 118 (1982) 359.
\item A. Sen and C. Vafa, \np 455 (1995) 165.
\item Y. Tanii, \pl 145 (1984) 197.
\item E. Cremmer, in {\it Superspace and Supergravity}, Eds. S. W. Hawking
and Rocek (Cambridge University Press), 1981.

\end{enumerate}
\end{document}